\documentclass[11pt,prd,preprint,nofootinbib,superscriptaddress]{revtex4}

\usepackage{amsmath}
\usepackage{amsfonts}
\usepackage{graphicx}
\usepackage{color}

\mathchardef\mhyphen="2D

\newcommand\ft[2]{{\textstyle\frac{#1}{#2}}}
\newcommand\nn{{\nonumber}}
\newcommand{\nc}{\newcommand}
\newcommand{\Li}{\text{Li}}

\nc{\bea}{\begin{eqnarray}} \nc{\eea}{\end{eqnarray}} \nc{\be}{\begin{equation}} \nc{\ee}{\end{equation}} \nc{\barr}{\begin{array}}
\nc{\earr}{\end{array}}
\nc{\btop}[2]{\genfrac{[}{]}{0pt}{}{\,#1\,}{\,#2\,}}
\nc{\zs}{z_\star}
\nc{\zsbar}{\bar{z}_\star}
\nc{\Pol}{{\text{Pol}(\Delta)}}
\nc{\Polk}{{\text{Pol}_k(\Delta)}}
\nc{\Poltk}{{\widetilde{\text{Pol}}_k(\Delta)}}
\nc{\Dd}{\mathcal{D}}
\nc{\DD}{{\Delta^{\frac{1}{2}}}}
\nc{\dE}{{}_{\partial E}}
\nc{\dEtw}{{}_{\partial E^2}}
\nc{\dEth}{{}_{\partial E^3}}
\nc{\dEfo}{{}_{\partial E^4}}
\nc{\sub}{\text{sub}}
\nc{\dive}{\text{div}}
\nc{\QNM}{\text{QNM}}
\nc{\y}{\frac{2\pi\hat m}{\sqrt{X}}}
\nc{\tk}{{\mathfrak t}}
\nc{\gtw}{{g_{(2)}}}
\nc{\Rtw}{{R_{(2)}}}
\nc{\Rf}{{R_{(\text{eff})}}}
\nc{\gf}{{g_{(\text{eff})}}}
\nc{\df}{d_{\text{eff}}}
\nc{\nbtw}{\nabla^2_{(2)}}
\nc{\nbf}{{\nabla_{(\text{eff})}}}
\nc{\ch}{{\text{cosh}}}
\nc{\sh}{{\text{sinh}}}

\begin{document}

\title{Computing black hole partition functions from quasinormal modes}

\author{Peter Arnold}
\email{parnold@virginia.edu}
\affiliation{Department of Physics, University of Virginia,\\
Box 400714, Charlottesville, VA 22904, USA}
\author{Phillip Szepietowski}
\email{p.g.szepietowski@uu.nl}
\affiliation{Institute for Theoretical Physics and Center for Extreme Matter and Emergent Phenomena,\\ Utrecht University, Leuvenlaan 4, 3584 CE Utrecht, The Netherlands}
\author{Diana Vaman}
\email{dv3h@virginia.edu}
\affiliation{Department of Physics, University of Virginia,\\
Box 400714, Charlottesville, VA 22904, USA}
\affiliation{Department of Physics, College of William and Mary, \\ Williamsburg, VA 23187-8795}

\begin{abstract}

We propose a method of computing one-loop determinants in black hole spacetimes (with emphasis on asymptotically anti-de Sitter black holes) that may be used for numerics when completely-analytic results are unattainable.  The method utilizes the expression for one-loop determinants in terms of quasinormal frequencies determined by Denef, Hartnoll and Sachdev in \cite{Denef:2009kn}. 
A numerical evaluation must face the fact that the sum over the quasinormal modes, indexed by momentum and overtone numbers, is divergent. 
A necessary ingredient is then a regularization scheme to handle the divergent contributions of individual fixed-momentum sectors to the partition function. To this end, we formulate an effective two-dimensional problem in which a natural refinement of standard heat kernel techniques can be used to account for contributions to the partition function at fixed momentum. We test our method in a concrete case by reproducing the scalar one-loop determinant in the BTZ black hole background. We then discuss the application of such techniques to more complicated spacetimes.

\end{abstract}

\maketitle

%%%%%%%%%%%%%%%%%%%%%%%%%%%%%%%%%%%%%%%%%%%%%%%%%%%%%%%%%%%%%%%%%%%%%%%%%%%%%

\section{Introduction}

With the advent of gauge-gravity duality, the study of classical fields in non-trivial gravitational spacetimes, and in particular in Anti-de Sitter (AdS) space, has received an incredible amount of attention. It is then natural to ask, what are the effects of quantum fluctuations around the classical gravitational saddle point? One obvious question along these lines is to consider the semi-classical calculation of the quantum gravity partition function, the study of which has a long history \cite{Gibbons:1994cg}. In the context of gauge-gravity duality, the gravity partition function in asymptotically AdS spacetimes is equated with the partition function of a strongly coupled conformal field theory (CFT) in the ``large-$N$" limit. One-loop contributions to the quantum gravity partition function then correspond to $``1/N"$ corrections to the partition function of the boundary field theory.%
\footnote{
  In the most familiar example, $N$ refers to the rank of an
  $\mathcal N = 4$ supersymmetric $SU(N)$ gauge theory, and the
  leading ``$1/N$'' correction scales as $1/N^2$.
  There are known stringy
  corrections of order $\lambda^{1/2}/N^2$ (where $\lambda$ is the
  't Hooft coupling) \cite{GKT,MPS} in addition to
  $\lambda^0/N^2$ one-loop gravity corrections that are the type of
  correction explored in the current paper.
}

A particularly interesting application of gauge-gravity duality is in the study of strongly coupled large-$N$ gauge theories at finite temperature. In the duality such systems are described by an asymptotically anti-de Sitter\footnote{We will focus on Schwarzschild anti-de Sitter black holes, in which case the dual gauge theory is a conformal field theory at finite temperature.} (AdS) black hole. One-loop corrections in such a black hole background then give a window into finite-$N$ corrections to thermodynamic and transport properties of the gauge theory plasma. Such finite-$N$ corrections are of interest as there are phenomena in the field theory which simply cannot be seen in the strict large-$N$ limit. For example, hydrodynamic long-time tails are not visible in classical gravity at infinite-$N$ \cite{Kovtun:2003vj} but manifest as a one-loop correction in the bulk \cite{CaronHuot:2009iq}. Other interesting examples include quantum oscillations in the presence of a magnetic field \cite{Denef:2009yy,Hartnoll:2009kk}, restoration of the Coleman-Mermin-Wagner theorem \cite{Anninos:2010sq}, non-Fermi liquid response \cite{Faulkner:2010da,Faulkner:2013bna} and quantum electron stars \cite{Hartnoll:2010gu,Allais:2012ye,Allais:2013lha}.

The computation of one-loop partition functions in black hole spacetimes is notoriously difficult. In \cite{Denef:2009kn}, Denef, Hartnoll and Sachdev (DHS) gave a beautiful expression for the one-loop determinant of a bulk field. The result of \cite{Denef:2009kn} expresses the one-loop determinant as a very explicit function in terms of a sum of the quasinormal frequencies of the bulk fluctuation. This function uniquely specifies the temperature dependence of the one-loop determinant, up to a set of ultra-violet (UV) local terms that can be computed in an asymptotic expansion. \cite{Denef:2009kn} provides several examples where the quasinormal mode spectrum can be computed analytically and then used to compare their formula with known results in simple cases (see also \cite{Datta:2011za,Datta:2012gc,Zhang:2012kya,Zojer:2012rj,Keeler:2014hba,Keeler:2016wko,Maloney:2016gsg}).

One drawback of applying the method of \cite{Denef:2009kn} is that for most black hole spacetimes one does not have an analytic expression for the quasinormal mode spectrum. Instead one typically computes the spectrum numerically. This poses a difficulty if one would like to compute the one-loop determinant using the results of \cite{Denef:2009kn}. The difficulty lies in the fact that the quasinormal mode sum which computes the determinant is UV divergent. In practice, these UV divergences manifest themselves in two ways. The quasinormal mode spectrum depends on two types of quantum numbers. The first labels the momentum transverse to the radial direction of the black hole. And the second is a quantum number associated with the radial direction, which in the Euclidean AdS black hole is the normal mode number associated with requiring normalizable conditions at the asymptotic boundary. UV divergences of the one-loop determinant occur when either of these quantum numbers becomes large. A consistent and pragmatic method of regularizing these divergences is the main goal of this paper.

Our goal is to extend the formalism of \cite{Denef:2009kn} to include spacetimes for which the quasinormal mode spectrum is not known analytically. We will describe a method which uses asymptotic WKB expressions of the quasinormal frequencies to effectively regulate the large radial momentum divergences. For the large transverse momentum divergence, we develop a new heat kernel expansion which is valid for both fixed and asymptotically-large transverse momenta.  This expansion, together with the details of organizing the calculation of the one-loop determinant to separate analytically-computable divergences from finite contributions (which may be computed numerically), are the primary technical contributions of this paper. As such, many results of the derivation of the heat kernel (although tedious) are included in Appendix \ref{app:HK}.

This paper is organized as follows. We begin with some preliminaries in section \ref{sec:Prelim} to introduce notation. In section \ref{sec:oneloop} we briefly recap the results of \cite{Denef:2009kn} and outline our numerical procedure for computing the determinant. Section \ref{sec:regfixedk} is devoted to understanding the UV asymptotics of one-loop determinants at fixed values of quantum numbers such as the momentum along the spatial boundary directions. In section \ref{sec:BTZ} we provide an explicit numerical calculation of the one-loop determinant in the three dimensional BTZ black hole. Comparison of our result with known analytic results in this case provides a modest proof of principle of our numerical procedure of computing one-loop determinants. Finally, in section \ref{sec:disc} we conclude with a discussion of future directions and potential caveats of applying our prescription in more complicated scenarios.

% ============================================================================

\section{Preliminaries}\label{sec:Prelim}

In order to set up our discussion of one-loop determinants we first set conventions and detail the types of background geometry and fluctuations that we will analyze. We will study asymptotically-AdS black holes/branes. Our primary example is the AdS Schwarzschild black hole\footnote{The black brane solution is given by the same metric (\ref{eq:AdSSMetric}) with $f(r) = \frac{r^2}{L^2}(1-\frac{r_h^d}{r^d})$ and $d\Omega_{d-1}^2$ replaced by the (normalized) flat metric $\frac{d\vec{x}^2}{L^2}$.} with metric given by
\begin{equation}\label{eq:AdSSMetric}
ds^2 = -f(r)\, dt^2 + \frac{dr^2}{f(r)} + r^2 d\Omega_{d-1}^2,
\end{equation}
where
\begin{equation}\label{eq:f}
f(r) = 1 - \frac{M}{r^{d-2}} + \frac{r^2}{L^2}
\end{equation}
and $d\Omega_{d-1}^2$ is the metric on $\mathbb S^{d-1}.$ We refer to the space transverse to the $r$-$t$ plane as the transverse space. We will be interested in one-loop corrections to the Euclidean partition function at temperature $T$. It is natural to Wick rotate to periodic time as $t = -i \tau$
where $\tau$ has period given by the inverse temperature, so that $\tau \sim \tau + 1/T.$

The principal example in this paper will be scalar fluctuations about this background. The Laplacian acting on a scalar $\phi$ is
\begin{eqnarray}\label{eq:AdSSscalarLaplacian}
\nabla^2 \phi &=& \frac{1}{\sqrt{g}} \partial_\mu (\sqrt{g} g^{\mu\nu} \partial_\nu \phi)\nn\\
&=& \left[\frac{1}{r^{d-1}}\partial_r(r^{d-1} f(r) \partial_r) + \frac{1}{f(r)} \partial_\tau^2  + \frac{1}{r^2}\nabla^2_{\Omega}\right]\phi.
\end{eqnarray}
A massive scalar will then satisfy the equation of motion
\begin{equation}
-\nabla^2\phi + m^2\phi = 0.
\end{equation}
In the context of holography, for asymptotically AdS$_{d+1}$ spacetimes the mass $m$ is related to the conformal dimension $\Delta$ of dual operators. For scalar operators this relation is simply $m^2L^2 = \Delta(\Delta-d).$

% ============================================================================

\section{One-loop determinants}\label{sec:oneloop}

In this section we present a method to compute one-loop determinants for fluctuations about static spacetimes and, in particular, about asymptotically-AdS black holes and black branes. We will begin with an overview of the results in \cite{Denef:2009kn}, which provide a method of computing determinants using the quasinormal mode fluctuations about the background geometry. We will then describe a proposal for extending these results to examples in which the quasinormal modes are only known numerically. In particular, for the cases of interest, the one-loop determinant can be separated into contributions from fixed-momentum\footnote{We will use the term momentum to refer to the quantum numbers of eigenmodes on the transverse space. For the case of black branes these correspond to continuous momenta along the transverse directions, whereas for black holes they label the eigenvalues of spherical harmonics.} sectors. The main obstacle in applying the formalism of \cite{Denef:2009kn} in such situations is that each fixed-momentum determinant is divergent and, furthermore, the subsequent sum over momenta is also divergent. We will see that the first of these divergences can be handled by an appropriate WKB analysis of the quasinormal modes at fixed momentum $k$, while the second divergence will be addressed later, in section \ref{sec:regfixedk}.

% ----------------------------------------------------------------------------

\subsection{One-loop determinants and quasinormal modes}

To begin, let us discuss the DHS formalism \cite{Denef:2009kn}.
In asymptotically AdS spacetimes, the quasinormal mode spectrum of fluctuations provides a natural basis for linearized perturbations about a background spacetime.\footnote{It should be emphasized that in a Lorentzian context the quasinormal modes do not form a complete basis for arbitrary solutions of the Laplacian. In terms of discussing the contributions to the Euclidean path integral we require analyticity of solutions to the Laplacian in imaginary time. In this case the quasinormal modes are related to normal modes which can form a complete basis of such solutions \cite{Warnick:2013hba}.} From the holographic point of view, the quasinormal modes determine the poles of the retarded Greens function of the operator dual to the field in question. In this sense, the quasinormal modes provide the closest thing to quasi-particle-like excitations in the strongly coupled dual field theory.

The key insight of DHS \cite{Denef:2009kn} is to realize that the quasinormal mode spectrum of an operator also determines the poles of the corresponding one-loop determinant.\footnote{We will use the terms partition function and one-loop determinant interchangeably. For a bosonic operator, the one-loop determinant appears in the denominator of the partition function and zero modes of the differential operator correspond to poles of the partition function. For fermions, the determinant appears in the numerator and zero modes correspond to zeroes of the partition function.} This can be seen by treating the partition function of an operator as a meromorphic function of its conformal dimension, $\Delta,$ which (for scalar fields) is related to the mass of the perturbation by $m^2L^2 = \Delta(\Delta-d)$. The poles of the one-loop determinant correspond to zero modes of the differential operator in Euclidean signature. Normalizability of a zero mode near the boundary ($r\rightarrow \infty$) fixes the behavior of the solution in terms of $\Delta.$ Matching this behavior with regularity of the zero modes at the origin of the Euclidean black hole then relates the values of $\Delta$ to the thermal frequencies $\omega_n = 2\pi n T,$ for integral $n$. When this relation is Wick rotated back to Lorentzian signature it becomes identical to the condition for the existence of a quasinormal mode. Therefore, as a complex function of $\Delta$, the poles of the partition function will occur precisely when $\Delta$ is such that a quasinormal mode (as a function of $\Delta$) coincides with a Wick rotated thermal frequency. If we denote the quasinormal frequencies by $z_\star(\Delta),$ this means that poles of the one-loop determinant occur at complex values of $\Delta$ such that
\begin{equation}
z_\star(\Delta) = i \omega_n = 2\pi i T n.
\end{equation}
As described in \cite{Coleman}, given a meromorphic function of $\Delta$ that has the correct poles one can determine the entire function by taking the limit $\Delta\rightarrow\infty$ and matching to an appropriate asymptotic of the function. As long as one can compute the large $\Delta$ asymptotics of the determinant, using for example the heat kernel, one can use this procedure to completely determine the one-loop determinant in terms of the quasinormal mode frequencies.

Assuming the meromorphicity properties described in the previous paragraph, DHS \cite{Denef:2009kn} proposed the following form for the partition function of a complex scalar\footnote{\cite{Denef:2009kn} also determines the form of the determinant for arbitrary bosonic and fermionic operators. For simplicity we will focus on scalar operators in the present work.} field:
\begin{equation}\label{eq:DHS}
Z = e^{\Pol} \prod_{\zs,\,\zsbar}\frac{\sqrt{\zs \zsbar}}{4\pi^2 T}\,\Gamma{\left(\frac{i\zs}{2\pi T}\right)}\,\Gamma{\left(\frac{-i\zsbar}{2\pi T}\right)},
\end{equation}
where $\zs$ ($\zsbar$) are the quasinormal frequencies with ingoing (outgoing) boundary conditions at the horizon and $T$ is the Hawking temperature of the background. The function $\Pol$ is a polynomial of $\Delta$ which is to be determined by matching to a large mass (large $\Delta$) expansion of the partition function. The function (\ref{eq:DHS}) is constructed such that it has poles whenever a quasinormal mode frequency $\zs(\Delta)$ coincides with a Wick rotated normal mode frequency $\omega = 2\pi i T n.$

% ----------------------------------------------------------------------------

\subsection{Fixed-momentum determinants}

In most non-extremal, finite temperature geometries the quasinormal mode spectrum is not analytically known and one has to resort to numerical methods. One expands the fluctuations in eigenfunctions of the transverse Laplacian and numerically determines the quasinormal mode spectrum. The end result is a spectrum of frequencies at a fixed value of the transverse momentum quantum number. In the case of a flat boundary geometry the transverse fluctuations are plane waves (for spherical transverse spaces these are spherical harmonics) and the state is labeled by the momentum $k$. For brevity, we will refer to quantities at fixed transverse quantum number as being at fixed $k$, even when referring to non-flat boundary geometries.

In order to compute the one-loop determinant one must sum over the spectrum at fixed $k$, and then later perform a sum over the momentum eigenvalues $k$. The sum over the fixed-$k$ quasinormal mode spectrum will be divergent. However, as long as one can determine the large frequency asymptotics (say in a WKB approximation), this divergence can be subtracted to yield a finite sum.

To make this discussion precise, factorize the partition function into fixed momentum sectors, writing
\begin{equation}\label {eq:Zprod}
Z = \prod_k Z_k
\end{equation}
where $Z_k$ is the fixed-$k$ partition function. Equation (\ref{eq:DHS}) can be written in this form, with $Z_k$ given by
\begin{equation}\label{eq:fixedkZ}
Z_k =  e^{\Polk} \prod_{\zs(k),\,\zsbar(k)}\frac{\sqrt{\zs(k) \zsbar(k)}}{4\pi^2 T}\,\Gamma{\left(\frac{i\zs(k)}{2\pi T}\right)}\,\Gamma{\left(\frac{-i\zsbar(k)}{2\pi T}\right)},
\end{equation}
or equivalently
\begin{equation}\label{eq:logZk}
\ln Z_k =  \Polk + \sum_{\zs(k),\,\zsbar(k)}\ln\left[\frac{\sqrt{\zs(k) \zsbar(k)}}{4\pi^2 T}\,\Gamma{\left(\frac{i\zs(k)}{2\pi T}\right)}\,\Gamma{\left(\frac{-i\zsbar(k)}{2\pi T}\right)}\right].
\end{equation}
The only difference with (\ref{eq:DHS}) is that (i) the quasinormal modes in the sum are restricted to the values at fixed $k$ and (ii) the exponential prefactor now contains a polynomial of $\Delta$ with $k$-dependent coefficients.

The quasinormal mode sum in (\ref{eq:logZk}) is divergent, which would be
problematic for a numerical calculation.
Our general strategy will be to find a good analytic approximation to
the {\it divergent}\/ piece of the sum (which can then be regulated)
and to only use numerics for the convergent piece that remains.
The divergence of (\ref{eq:logZk}) comes from arbitrarily large
quasinormal mode frequencies.  For those frequencies, one may
generically use the WKB expansion (instead of numerics) to determine
the frequencies.
In the WKB expansion, the quasinormal mode frequencies are labeled by an integer mode number $n \geq 0$ such that the quasinormal modes have an expansion of the schematic form\footnote{This is the form for scalar fields in asymptotically-AdS black holes \cite{Natario:2004jd,Musiri:2005ev}. Other bosonic fluctuations presumably have a similar structure, although the $\Delta$ and $k$-dependence of such an expansion for arbitrary spin fields has not been worked out. In addition, there are known cases, such as fermionic fields in $d>2$ and gauge fields in $d=3$, where this expansion develops $\ln n$ terms which include logarithmic dependence on functions of $k$ and $\Delta$ \cite{Musiri:2005ev,Arnold:2013gka,Arnold:2013zva}.}
\begin{eqnarray}\label{eq:WKBQNM}
\frac{z_{\star,n}(k)}{2\pi T} &\simeq& A\bigl[ n + B(\Delta,k) + C(\Delta,k)\,n^{-\delta} + \cdots\bigr]
\end{eqnarray}
where $A$ is a dimension dependent complex constant, $B(\Delta,k)$ and $C(\Delta,k)$ are complex functions of $\Delta$ and $k,$ and $\delta$ is a positive dimension-dependent number. For scalar fields in AdS-Schwarzschild the functions $B(\Delta,k)$ and $C(\Delta,k)$ are polynomial in $\Delta$ and $k$. In particular for AdS-Schwarzschild black branes in $d>2$, one finds the leading terms in the expansion to be
\begin{equation}
B(\Delta,k) = \frac{\Delta}{2} - \frac{1}{2} - i \frac{\ln 2}{2\pi}, \qquad
C(\Delta,k) \propto k^2, \qquad \delta = \frac{d-2}{d-1} 
\end{equation}
for scalar fields
(the case we will focus on).  The $``\cdots"$ in (\ref{eq:WKBQNM}) represents terms with higher negative powers of $n$, which can be systematically determined in this expansion.  For scalars in AdS-Schwarzschild, the coefficients of such terms will also be polynomial in $\Delta$ and $k$. 

We should note that for the BTZ black hole in $d=2,$ which is our test case in section \ref{sec:BTZ}, the expansion (\ref{eq:WKBQNM}) actually terminates such that $C(\Delta,k)$ and the $``\cdots"$s in (\ref{eq:WKBQNM}) all vanish. In fact, the exact quasinormal mode frequencies are known for arbitrary spin fields in the BTZ black hole background. In order to extend our results to higher dimensional black holes one must compute the expansion (\ref{eq:WKBQNM}) at least to high enough order in $1/n$ to remove all divergences in the sum over $n$ in the fixed-$k$ partition function. This would require employing techniques such as those in \cite{Musiri:2005ev,Siopsis:2008xz} to compute the asymptotic quasinormal spectrum to higher order in perturbation theory.  We hope to return to this in the near future.

Once the $z_{*,n}(k)$ are known to sufficiently high order in $1/n$ one can compute the divergent terms in (\ref{eq:fixedkZ}) and explicitly subtract them off. Doing so we can define a subtracted sum for the logarithm of the partition function,
\begin{equation}\label{eq:Zksub}
\ln Z_k^{\sub} = \ln Z^{\QNM}_k - (\ln Z^{\QNM}_k)_{\dive} .
\end{equation}
Above, $Z^{\QNM}_k$ (QNM for ``quasinormal mode'') refers to the original
divergent sum in (\ref{eq:logZk}) without the $\Polk$ term (to which we
return shortly),
\begin{equation}\label{eq:logZkQNM}
\ln Z^{\QNM}_k =  \sum_{\zs(k),\,\zsbar(k)}\ln\left[\frac{\sqrt{\zs(k) \zsbar(k)}}{4\pi^2 T}\Gamma{\left(\frac{i\zs(k)}{2\pi T}\right)}\Gamma{\left(\frac{-i\zsbar(k)}{2\pi T}\right)}\right].
\end{equation}
$(\ln Z^{\QNM}_k)_{\dive}$ is defined as the asymptotic (large $n$) WKB
expansion of $\ln Z^{\QNM}_k$, {\it truncated}\/ at a finite order that includes
all terms that diverge when summed as in (\ref{eq:logZkQNM}).
The superscript ``sub'' on $\ln Z_k^{\sub}$ stands for ``subtracted.''

The expression (\ref{eq:Zksub}) is, by construction, a finite sum over $n$.  It will differ from (\ref{eq:logZk}) by $k$-dependent polynomial terms in $\Delta$ which can be absorbed into $\Polk$.  We call the new polynomial $\Poltk$. The log of the full fixed-$k$ partition function can then be written
\begin{equation}\label{eq:ZksubwPol}
\ln Z_k = \Poltk + \ln Z_k^{\sub}.
\end{equation}
We will operate under the assumption that this factorization is possible---in particular that the fixed-$k$ partition function satisfies the same analyticity properties as the full partition function, so that $\Poltk$ can be determined from a local expression in the $r$-$\tau$ plane transverse to the spatial boundary directions and can be calculated in the $\Delta\rightarrow \infty$ limit.

In the next section we will describe how one can use a modified heat kernel to determine the ultraviolet (UV) large-$\Delta$ asymptotics at fixed-$k$ in order to determine $\Poltk$ and, furthermore, how to utilize this heat kernel to regulate the sum over momentum states.

% ============================================================================

\section{Regularization and a fixed-\boldmath$k$ heat kernel}\label{sec:regfixedk}

We now move on to the discussion of regularizing the fixed-$k$ partition function described in the previous section. There are two issues with the fixed-$k$ partition function as expressed in (\ref{eq:fixedkZ}), even after subtracting out the large frequency asymptotics as in (\ref{eq:Zksub}). First, it is divergent as a product over $k.$ This requires a method of determining the large-$k$ asymptotics of the fixed-$k$ partition function and consistently subtracting the divergent contributions to (\ref{eq:fixedkZ}) when summed over all $k$. Second, in order to determine $\Poltk$ we will need a way of determining the large-$\Delta$ asymptotics of the fixed-$k$ partition function. We will find that both of these issues can be taken care of with an appropriate fixed-$k$ heat kernel. The goal of this section is to construct this fixed-$k$ heat kernel.

% ----------------------------------------------------------------------------

\subsection{Reducing the Laplacian to a two-dimensional problem}

To derive the form of the fixed-$k$ heat kernel it is convenient to rewrite the Laplacian as an effective two-dimensional operator, where the $k$ dependence is explicitly packaged into a potential term as opposed to arising as a quantum number due to the background geometry.

Concretely, consider again the scalar Laplacian in the AdS-Schwarzschild black hole (\ref{eq:AdSSscalarLaplacian}). We can expand in eigenmodes of the transverse Laplacian.  These satisfy
\begin{equation}
\nabla^2_{\Omega_{d-1}} \varphi_k(x_\perp) = - k^2 \varphi_k(x_\perp),
\end{equation}
where $k^2$ labels the eigenvalues of the transverse Laplacian. In particular, $k^2$ is dimensionless and given by $k^2 = p^2 L^2$ and $k^2 = l(l+d-2)$, with $l$ a non-negative integer, for flat and spherical boundaries, respectively. Expanding in these modes schematically as
\begin{equation}
\phi(r,\tau,x_\perp) = \sum_k \phi_k(r,\tau) \, \varphi_k(x_\perp)
\end{equation}
the Laplacian acting on the modes $\phi_k$ becomes
\begin{equation}\label{eq:fixedkLaplacian}
\nabla^2 \phi_k = \left[\frac{1}{r^{d-1}}\partial_r(r^{d-1} f(r) \partial_r) + \frac{1}{f(r)} \partial_\tau^2  - \frac{k^2}{r^2}\right]\phi_k.
\end{equation}
It is natural to rescale
$\phi_k(r,\tau) = \left(\frac{L}{r}\right)^{(d-1)/2} \psi_k(r,\tau).$
In terms of $\psi_k,$ the Laplacian is
\begin{eqnarray}
\nabla^2 \psi_k(r,\tau) &=& \left[\partial_r\left(f(r)\partial_r\right) + \frac{1}{f(r)}\partial_\tau^2 - \frac{k^2}{r^2} - \frac{(d-3)(d-1)f(r)}{4r^2} - \frac{2(d-1)f'(r)}{4r}\right]\psi_k(r,\tau).
\end{eqnarray}
We can rewrite this as
\begin{eqnarray}\label{eq:2DLaplace}
\nabla^2 \psi_k &=& \left[\nbtw - U(r)\right]\psi_k,
\end{eqnarray}
where $\nbtw$ is the Laplacian for a scalar in the two-dimensional background
\begin{equation}\label{eq:2Dslice}
ds_{(2)}^2 = f(r)\, d\tau^2 + \frac{dr^2}{f(r)}.
\end{equation}
Here, $f(r)$ is given in (\ref{eq:f}) and we have defined the potential
\begin{equation}\label{eq:U}
U(r) = \frac{k^2}{r^2} + \frac{(d-3)(d-1)f(r)}{4r^2} + \frac{2(d-1)f'(r)}{4r}.
\end{equation}
For later convenience,  we also quote the value for the Ricci curvature
\begin{equation}\label{eq:Ricci}
\Rtw_{ab} = -\frac{1}{2}f''(r)g_{ab}.
\end{equation}
of the two-dimensional metric (\ref{eq:2Dslice}).
Note that the geometry (\ref{eq:2Dslice}) is just the naive dimensional reduction of the original geometry (\ref{eq:AdSSMetric}). This represents the effective geometry, along with the potential (\ref{eq:U}), that each fixed-$k$ mode function probes. Here we use it as a construct so that we can apply standard heat kernel techniques to determine the asymptotics of the fixed-$k$ partition function. We therefore re-interpret the fixed-$k$ partition function $Z_k$ for a scalar in the AdS black hole/brane spacetime as the partition function of a scalar in the two dimensional geometry (\ref{eq:2Dslice}) with the potential (\ref{eq:U}).

% ----------------------------------------------------------------------------

\subsection{The Heat Kernel}

A very useful method of determining the UV asymptotics of one-loop determinants is to compute the heat kernel associated with the differential operator. (For a comprehensive review of heat kernel techniques, see \cite{Vassilevich:2003xt}.) Considering a generic two-derivative operator $D,$ one constructs the heat kernel as the solution $K(x,x';\tk)$ of
\begin{equation}
(\partial_{\tk} + D + m^2)\, K(x,x';\tk) = 0,
\end{equation}
where we take $D$ to act on the variable $x$ and impose the boundary condition $K(x,x';0) = \delta^{(d+1)}(x,x').$

Given a solution $K(x,x';\tk),$ the logarithm of the one-loop determinant is determined as
\begin{equation}\label{eq:detD}
\ln \det(D+m^2) = \text{const} - \int d^{d+1}x \sqrt{g}\int_0^\infty \frac{d\tk}{\tk}\> K(x,x;\tk),
\end{equation}
where ``const'' corresponds to an undetermined overall normalization of the partition function. It is possible to solve the heat kernel in a small-$\tk$ expansion, which we refer to as the heat kernel expansion. This is particularly useful in determining the high energy asymptotics of the heat kernel, and hence, also of the partition function. The heat kernel expansion gives the following expression for $K(x,x';\tk)$ in the $x\rightarrow x'$ coincidence limit,\footnote{In this equation we explicitly write the full dependence on $(x,x).$ Aside from Appendix \ref{app:HK}, in the rest of the paper we will always write expressions in the coincidence limit and will therefore suppress the second index.}
\begin{equation}\label{eq:HK}
K(x,x;\tk) = \,(4\pi\tk)^{-(d+1)/2}\sum_{j=0} a_{2j}(x)\, \tk^j e^{-\tk m^2},
\end{equation}
where the coefficients $a_{2j}(x)$ are local functions of the background geometry constructed out of curvature invariants. Taking the operator to be $D = -(\nabla^2 +E),$ with $\nabla^2$ the scalar Laplacian and $E$ an arbitrary potential, the first several heat kernel coefficients take the universal form \cite{Vassilevich:2003xt}
\begin{subequations}\label{eq:HKacoeff}
\begin{eqnarray}
a_0(x) &=& 1, \\
a_2(x) &=& \frac{1}{6}R + E, \\
\label{eq:HKacoeff4}
a_4(x) &=& \frac{1}{72}R^2 - \frac{1}{180}R_{\mu\nu}R^{\mu\nu} + \frac{1}{180} R_{\mu\nu\rho\sigma}R^{\mu\nu\rho\sigma} + \frac{1}{30}\nabla^2 R + \frac{1}{6}E_{;\mu}{}^\mu + \frac{1}{6}R E  + \frac{1}{2}E^2.
\end{eqnarray}
\end{subequations}
The above expansion is sufficient for discussing the UV asymptotics of the partition function for $d \le 4.$ In particular, if the UV contribution to the integral in (\ref{eq:detD}) is regulated by a strict cut-off $\tk > 1/\Lambda^2$ then for $d \le 4$ all divergences in the $\Lambda\rightarrow\infty$ limit are contained in the terms present above. These terms also suffice in determining the large mass (large $\Delta$) limit of the determinant. Precisely this type of regulator was used in \cite{Denef:2009kn} to determine $\Pol$ by matching the large $\Delta$ limits of the heat kernel and the logarithm of (\ref{eq:DHS}), completely fixing the normalization of the free energy (up to an overall $\Delta$-independent constant). For our purposes we will need a slightly refined version of the heat kernel, as we discuss in the following subsection.

In order to compare and contrast with the discussion we will have in
the next subsection, it is worth taking a moment to briefly review why
the heat kernel expansion can be used to study the large mass limit.
Combining (\ref{eq:detD}) and (\ref{eq:HK}) formally gives
\begin{equation}
\ln \det(D+m^2) = \text{const}
  - 
  (4\pi)^{-(d+1)/2} \int d^{d+1}x \> \sqrt{g} \sum_j a_{2j}(x) \int_0^\infty
    \frac{d\tk}{\tk} \> \tk^{j-(d+1)/2}e^{-\tk m^2} .
\end{equation}
For large enough $j$, the $\tk$ integral is dominated by $\tk \sim m^{-2}$
and so is of order $m^{-2j+d+1}$: the expansion in $j$
produces an expansion in $m^{-2}$.

% ----------------------------------------------------------------------------

\subsection{The Heat Kernel at Fixed \boldmath$k$}\label{sec:kHK}

In order to regulate the asymptotics of the fixed-$k$ partition function we will need an expression for the heat kernel which has the correct asymptotic behavior both at large $\Delta$ {\it and} large $k$.
In particular, the product (\ref{eq:Zprod}) over $k$ sectors gives\footnote{
	The $\sum_k$ and $\int_k$ forms in (\ref{eq:sumk}) assume that $\ln Z_k$ is
	normalized with discrete $k$ and continuum $k$ conventions respectively.
	In this section, we will treat $\sum_k$ and $\int d^{d-1}k/(2\pi)^{d-1}$
	interchangeably and leave the normalization implicit.  When we take
	up the BTZ black hole in section \ref{sec:BTZ}, $k$ will be discrete, and explicit
	formulas will use the corresponding normalization for $\ln Z_k$.
	In the appendices, we will occasionally discuss the black brane
	limit, where $k$ is continuous, but we will not bother to be explicit
	about changes to normalization factors that appear in switching
	between the discrete and continuum $k$ normalizations involving
	the size $\int d^{d-1}x$ of the space of
	transverse coordinates.}
\begin {equation}
  \ln Z = \sum_k \ln Z_k = \int \frac{d^{d-1}k}{(2\pi)^{d-1}} \> \ln Z_k .
\label {eq:sumk}
\end {equation}
Since numerics are not well suited to divergent expressions, we will need
to be able to subtract out all the contributions to $\ln Z_k$ that
give divergent contributions to the integral (or sum) over $k$.
For that, we will need to find the large-$k$ expansion of $\ln Z_k$
up to order $k^{-(d-1)}$.

Recall that our strategy for working at fixed $k$ is to interpret
the problem as a 2-dimensional\footnote{In the language of Appendix \ref{app:HK}, the effective dimension is $\df + 1 = 2.$} problem (\ref{eq:2DLaplace}--\ref{eq:2Dslice})
in $r$ and $\tau$. Correspondingly, the generic heat kernel expression
(\ref{eq:detD}) becomes
\begin{equation}\label{eq:HK2}
\ln Z_k = \tfrac12 \ln \det_k(-\nabla^2+m^2)
= \tfrac12 \ln \det(-\nabla_{(2)}^2+U+m^2)
= \frac{1}{2}
  \int d^2x \sqrt{\gtw}\int \frac{d\tk}{\tk} K_k(x;\tk)
\end{equation}
with $x=(r,\tau)$ here and
\begin{equation}\label{eq:kHKexp2}
K_k(x;\tk) = \frac{1}{4\pi\tk}  \sum_{j=0}^\infty a_{2j}(x) \tk^j e^{-\tk m^2}
\end{equation}
and the $E$ in expressions (\ref{eq:HKacoeff}) for the coefficients
corresponding to
\begin{equation}\label{eq:E}
E= - U(r) = - \frac{k^2}{r^2} - \frac{(d-3)(d-1)f(r)}{4r^2} - \frac{2(d-1)f'(r)}{4r}.
\end{equation}

However, to reproduce the correct behavior at large $k$, it is necessary to modify the standard heat kernel expansion. To see this, note that $E$ above
contains a term proportional to $k^2$.
The coefficient $a_{2n}(x)$ in the heat kernel expansion (\ref{eq:HKacoeff})
contains a term proportional to $E^n$, which in our application is therefore
proportional to $k^{2n}$.  Each subsequent order in the expansion will contain higher and higher powers of $k^2$, and so the usual heat kernel expansion (\ref{eq:HK}) breaks down in the large-$k$ limit. Fortunately, there is a natural workaround.

Consider again the generic heat kernel expansion of an operator of the form $D= -(\nabla^2 + E)$.  First note that the terms with bare powers of $E$ in the heat kernel expansion (\ref{eq:HK}--\ref{eq:HKacoeff}) appear to exponentiate to $e^{\tk E}$.
So let us reorganize the heat kernel expansion to include the factor
$e^{\tk E}$ explicitly:
\begin{equation}\label{eq:fixedkHKexp}
K_k(x;\tk) = \frac{1}{4\pi\tk} \sum_{j=0}^\infty b_{2j}(x)\,\tk^j e^{-\tk m^2 + \tk\,E(x)},
\end{equation}
with
\begin{subequations}\label{eq:fixedkHKcoeffs1}
\begin{align}
b_0(x) &= 1, \\
b_2(x) &= \frac{1}{6}\Rtw, \\
b_4(x) &= \frac{1}{72}\Rtw^2 - \frac{1}{180}\Rtw_{\mu\nu}\Rtw^{\mu\nu} + \frac{1}{180} \Rtw_{\mu\nu\rho\sigma}\Rtw^{\mu\nu\rho\sigma} + \frac{1}{30}\nabla_{2}^2 \Rtw + \frac{1}{6}E_{;\mu}{}^\mu
\label{eq:HKb4}
\end{align}
\end{subequations}
[where all of the quantities, such as curvature tensors and covariant derivatives, are defined with respect to the two-dimensional geometry (\ref{eq:2Dslice})]. Putting the exponential factor $e^{-\tk E}$ explicitly in the heat kernel removes the problematic $E^n$ terms in the heat kernel coefficients $b_{2n}(x)$ and also provides a suppression of the large-$k$ sector for each term in the reorganized expansion.

However, there remain terms proportional to derivatives of $E$ (and hence proportional to $k^2$) in the new heat kernel coefficients $b_j$ above.  These terms do not appear to exponentiate, and we might worry that they spoil the convergence of the heat kernel expansion at large $k$.  Fortunately they do not, but we will see that one must keep more terms of the reorganized heat kernel expansion than one might have expected.

Here's the issue.  Consider the case of large $k$ (for fixed $m$ and $r$).
The exponential factor in (\ref{eq:fixedkHKexp}) will
effectively restrict the $\tk$ integration of
(\ref{eq:HK2}) to $\tk \lesssim r^2/k^2$.
There are now two opposing effects as we go to higher and higher orders
$j$ in the expansion: (i) $\tk^j$ will give us more and more powers of
$k^{-2}$ while (ii) we may get derivatives of $E$ appearing in the associated
coefficients $a_{2j}$, and each such derivative of $E$ will give a power
of $k^2$.  As an example, the $\tk^{-1} \times b_4 \tk^2$ term (i.e.\ $j{=}2$ term)
in (\ref{eq:fixedkHKexp})
has a contribution of order $k^0$ because of the $E_{;\mu}{}^\mu$ term
in (\ref{eq:HKb4}), and this is the same size as the $\tk^{-1} \times b_2 \tk$
term (i.e.\ $j{=}1$ term) in (\ref{eq:fixedkHKexp}).
Fortunately, we find that the contributions from higher and higher
orders in the reorganized expansion do not remain this size: they slowly
decrease (by powers of $k^{-2}$) in steps.  As an example, consider the
case $d=4$, for which we would like to analytically
extract the large $k$ dependence
down to $k^{-(d-1)} = k^{-3}$ in order to isolate the divergences
in (\ref{eq:sumk}).  We find that all of these terms are accounted
for by (\ref{eq:fixedkHKcoeffs1}) supplemented by%
\footnote{Given the effective two-dimensional geometry, it is straightforward to evaluate the $b_{i}$'s in terms of $f(r), E(r)$  and their derivatives:
\[b_2(x)=-\tfrac 16 \frac{d^2 f}{dr^2},\qquad
b_4(x)=\tfrac{1}{60}\bigg((\frac{d^2 f}{dr^2})^2-2 \frac{df}{dr}\frac{d^3 f}{dr^3} -2 f\frac{d^4 f}{dr^4} \bigg)+\tfrac 16\bigg(\frac{df}{dr}\frac{dE}{dr}+f\frac{d^2 E}{dr^2}\bigg),\]
\[b_6(x)\big|_{\dE}=\tfrac{1}{60}\bigg(
-\frac{df}{dr}\frac{d^2 f}{dr^2}\frac{dE}{dr}-f\frac{d^3 f}{dr^3}\frac{dE}{dr}+5f(\frac{dE}{dr})^2
+f\frac{d^2f}{dr^2}\frac{d^2E}{dr^2}+2(\frac{df}{dr})^2\frac{d^2 E}{dr^2}
+4f\frac{df}{dr}\frac{d^3E}{dr^3}+f^2\frac{d^4 E}{dr^4}\bigg),\]
\[b_8(x)\big|_{\dEtw}=\bigg(\tfrac{1}{80}( \frac{df}{dr})^2+\tfrac {1}{90}f\frac{d^2 f}{dr^2}\bigg)(\frac{dE}{dr})^2+\tfrac {1}{40}f^2(\frac{d^2E}{dr^2})^2+\tfrac{11}{120}f\frac{df}{dr}\frac{dE}{dr}\frac{d^2 E}{dr^2}+\tfrac1{30}f^2\frac{dE}{dr}\frac{d^3E}{dr^2},\]
\[b_{10}(x)\big|_{\dEth}=\frac{1}{45}f\frac{df}{dr}(\frac{dE}{dr})^3+\tfrac{11}{360}f^2(\frac{dE}{dr})^2\frac{d^2 E}{dr^2},\qquad 
b_{12}(x)\big|_{\dEfo}=\tfrac{1}{288}f^2(\frac{dE}{dr})^4.
\]
}
\begin{subequations}\label{eq:fixedkHKcoeffs2}
\begin{align}
b_6(x)\big|_{\dE} &=  \frac{1}{90}\Rtw^{\mu\nu}E_{;\mu\nu} + \frac{1}{36}\Rtw E_{;\mu}{}^\mu + \frac{1}{30} \Rtw^{;\mu} E_{;\mu} + \frac{1}{60}E_{;\mu}{}^\mu{}_\nu{}^\nu + \frac{1}{12} E_{;\mu}E^{;\mu}, \displaybreak[0]\\
b_8(x)\big|_{\dEtw} &=\frac{1}{72}\Rtw E_{;\mu}E^{;\mu} + \frac{1}{72}(E_{;\mu}{}^\mu)^2 + \frac{1}{90}E_{;\mu\nu}E^{;\mu\nu} + \frac{1}{60}E^{;\mu}E_{;\mu\nu}{}^\nu + \frac{1}{60}E_{;\mu}E_{;\nu}{}^{\nu\mu}, \displaybreak[0]\\
b_{10}(x)\big|_{\dEth} &= \frac{1}{60}E_{;\mu\nu}E^{;\mu}E^{;\nu} + \frac{1}{72}E_{;\mu}{}^{\mu}E^{;\nu}E_{;\nu}, \displaybreak[0]\\
b_{12}(x)\big|_{\dEfo} &= \frac{1}{12\cdot 4!}(E^{;\mu}E_{;\mu})^2.
\end{align}
\end{subequations}
Details, based on a modified Seeley-DeWitt expansion,
are given in Appendix \ref{app:HK}.
The subscript $\partial E^n$ above is used to denote that, in that coefficient, we have kept terms with at least $n$ factors of (derivatives of) $E$ and have dropped terms that are lower order in $E$.  Note, for example, that we have kept terms in $b_6$
that contribute to $\tk^{-1} \times b_{2j} \tk^j$ (and so the heat kernel
expansion) at order $k^0$ and $k^{-2}$,
but we have not bothered to include
the non-$E$ terms, which contribute at order $k^{-4}$.

Here's an equivalent way of characterizing which terms need to be kept.
Think of the reorganized heat kernel expansion as an expansion in small
$\tk$ except considering $\tk k^2$ as fixed\footnote{When making power counting arguments, we will treat $r$ as fixed and will often use $\tk k^2$ as shorthand for the dimensionless quantity $\tk k^2/r^2$. We will separately discuss the issue of boundary regularization ($r{\to}0$) later, in section \ref{sec:BTZfixedk}.} [in order to account for the fact
that $\tk k^2$ can be as large as $O(1)$].
Using the notation $O(\tk^n_{\text{eff}})$ to denote terms of $O(\tk^n)$ multiplied by arbitrary powers of $\tk k^2$,
the $b_0(x)$ term in the sum in (\ref{eq:HK2}) is
$O(\tk_{\text{eff}}^0)$,
the $\{b_2(x),b_4(x),b_6(x)\}$ terms are $O(\tk_{\text{eff}}^1)$, and the
$\{b_8(x),b_{10}(x),b_{12}(x)\}$ terms are $O(\tk_{\text{eff}}^2)$. The important thing to note about this power counting is that (after the constant term) the degree of divergence in $\tk_{\text{eff}}$
jumps by one power for every three powers of $\tk$ using the naive power counting. This behavior is implied by the heat equation and is necessary for the consistency of the fixed-$k$ heat kernel expansion. The origin of this power counting pattern is discussed in more detail in Appendix \ref{app:HK}.
Eqs.\ (\ref{eq:fixedkHKcoeffs1}) and (\ref{eq:fixedkHKcoeffs2}) give all
the terms necessary for determining the divergence of the free energy
for $d\le4$. In higher dimensions $d>4$, one needs additional
terms in order to capture all of the large-$k$ divergences.

We include a detailed discussion of the consistency of the expansion
and a derivation of the appropriate heat kernel coefficients in
Appendix \ref{app:HK}.  As a cross check, we show in Appendix
\ref{subsec:HKRecov} that our fixed-$k$ expansion reproduces the standard heat
kernel expansion if one integrates the fixed-$k$ heat kernel
over $k$ before integrating over $\tk$.

Before moving on, we should note a possible danger in our power-counting
arguments above.  We have discussed the large-$k$ expansion for fixed
$r$.  However, when computing $\ln Z_k$ as in (\ref{eq:HK2}),
we will eventually need to integrate over $r$, including arbitrarily
large values of $r$ for a given $k$.  Could that cause trouble for
our use of the preceding large-$k$ (fixed $r$) expansion? 
We will later briefly discuss
in section \ref{sec:warning}
(in the context of a concrete example) how we can sidestep
this issue, followed by a more thorough discussion of the problem in
appendix \ref{app:largishr}.  For now, we blithely ignore it.

% ----------------------------------------------------------------------------

\subsection{Determining the complete determinant}\label{subsec:Detdet}

Having determined the fixed-$k$ heat kernel we are now in a position to detail the appropriate regularization procedure to compute the full one-loop determinant.

The full form of the determinant is given by summing equation (\ref{eq:ZksubwPol}) over all momentum modes. Depending on the geometry this sum is either an infinite sum over discrete modes or an integral over continuous momenta. For notational clarity we will denote this as an integral, appropriate for black brane geometries with translationally invariant horizons. The logarithm of the partition function is given by
\begin{equation}\label{eq:logZdivk}
\ln Z = \int  \frac{d^{d-1}k}{(2\pi)^{d-1}} \left(\Poltk + \ln Z_k^{\sub}\right).
\end{equation}
The term $\Poltk$ can be determined by taking the large $\Delta$ limit of this expression and matching to the large $\Delta$ limit of the heat kernel in equation (\ref{eq:fixedkHKexp}) of the previous subsection. Another crucial use of the heat kernel arises when one considers the integral over momentum. The integral in equation (\ref{eq:logZdivk}) is divergent in the UV. To regularize this we need to subtract out the divergences arising in the large-$k$ regime of the integral. For this we again use the heat kernel (\ref{eq:fixedkHKexp}), however, now without taking the large $\Delta$ limit. Since (\ref{eq:fixedkHKexp}) was constructed to contain all of the UV divergences associated with the integral over momenta it should be sufficient to cancel all such divergences in the momentum integral in (\ref{eq:logZdivk}). Formally we may add and subtract the heat kernel expression from the QNM sum.

Let $K_k^{\rm trunc}(x;\tk)$ represent the truncation of the fixed-$k$ heat
kernel expansion (\ref{eq:fixedkHKexp}) to
contain just those terms that will give divergences when integrated
over $k$ for a given dimension $d$.
For example, for $d{=}4$, $K_k^{\rm trunc}(x;\tk)$ would contain
all of the terms in (\ref{eq:fixedkHKcoeffs1}) and (\ref{eq:fixedkHKcoeffs2}).
Define $I$ to be
the result of fully integrating this truncated heat kernel expansion
(with appropriate regularization), i.e.
\begin{equation}\label{eq:IntHK}
I = \frac{1}{2}\int \frac{d^{d-1}k}{(2\pi)^{d-1}} \int dr\,d\tau \int \frac{d\tk}{\tk}\> K_k^{\rm trunc}(x;\tk) \equiv \int \frac{d^{d-1}k}{(2\pi)^{d-1}}\, F(k).
\end{equation}
One can then re-write $\ln Z$ by adding and subtracting the large-$k$ heat kernel representation of the partition function:
\begin{eqnarray}\label{eq:logZfinal}
\ln Z &=& \int \frac{d^{d-1}k}{(2\pi)^{d-1}} \left(\Poltk + \ln Z_k^{\sub}\right) \nn \\
&=& I + \int\frac{d^{d-1}k}{(2\pi)^{d-1}} \left(\Poltk + \ln Z_k^{\sub} - F(k)\right).
\end{eqnarray}
The integral on the first line is the bare quasinormal mode representation and is divergent. On the second line we have added and subtracted the result from the large-$k$ heat kernel. The integrand in parentheses on the second line then gives a finite result when integrated over $k$ and can be computed numerically.

Note that while $I$ and $\int_k F(k)$ are formally equivalent, in practice both are infinite, and we will need to take care to consistently regularize our calculations of the different terms in (\ref{eq:logZfinal}).  Let $\tilde\Lambda$ be the momentum scale for UV regularization.%
\footnote{
  In this generic discussion, we will be a little bit sloppy and think of
  the UV cutoff as directly a cutoff $k \lesssim \tilde\Lambda$ on $k$.
  In the specific example of the next section, however, $\Lambda$ will be
  the usual cutoff used in heat kernel regularization, which we will see
  corresponds to a cutoff $\tilde\Lambda \sim r\Lambda$ on $k$.
  Also, when we refer to UV regularization in this paper, we are referring
  to the UV of the gravity theory.  In particular, we are not referring
  to boundary regularization of the asymptotically AdS space-time, which
  we will handle separately.
}
As we will see explicitly in the example of the next section, the UV divergences of $\Poltk$ and $F(k)$ cancel each other in the last line of (\ref{eq:logZfinal}), which is why we can do that $k$ integral numerically.  As a result, when separately deriving the divergent $\Poltk$ and $F(k)$ terms to use in that integrand, it is adequate to consider the limit of $k \ll \tilde\Lambda$, since the contribution from $k \sim \tilde\Lambda$ will disappear as $\tilde\Lambda \to \infty$.  In contrast, the $k$ integral (\ref{eq:IntHK}) defining $I$ is divergent.  So, when computing $I$, we must also correctly treat the $k \sim \tilde\Lambda$ case:
a $k \ll \tilde\Lambda$ approximation to $F(k)$ in (\ref{eq:IntHK}) will not do.
In appendix \ref{app:HK}, we show that the calculation of the integral $I$
yields the usual heat kernel result for the partition function (up to computable finite contributions for the case of compact horizons).

We now turn to a specific example to detail how this procedure works in
practice.

% ============================================================================

\section{Example --- BTZ black hole}\label{sec:BTZ}

We now turn to an application of the formalism described in the previous sections. In particular, we will use our method to compute the one-loop determinant of a scalar field in the BTZ black hole background. The partition function of a scalar field in BTZ had been previously computed using other methods in \cite{Mann:1996ze}. In fact, for this case, the quasinormal modes are known analytically and an exact result for the determinant was derived in \cite{Denef:2009kn}. Here this example will serve as a simple test case to illustrate the formalism developed in the previous sections.

The Euclidean BTZ black hole metric is given by
\begin{equation}\label{eq:BTZmet}
ds^2 = f(r) \, d\tau^2 + \frac{dr^2}{f(r)} + r^2 d\phi^2.
\end{equation}
This is of the form (\ref{eq:AdSSMetric}) with $d=2$ and $f(r)= \frac{r^2}{L^2}(1-\frac{r_h^2}{r^2}).$ The horizon radius is related to the temperature of the spacetime by $r_h = 2\pi T L^2.$ The coordinate $\phi$ can be chosen to be periodic with $\phi \sim \phi + 2\pi.$ One may also choose $\phi$ to not be periodic, in which case the metric (\ref{eq:BTZmet}) is a black brane instead of a black hole. In holography, periodic $\phi$ corresponds to placing the dual CFT on a spatial circle, whereas for non-periodic $\phi$ the dual CFT is defined on the real line. We will assume periodicity in $\phi$ in what follows; so we consider the black hole, but we will comment on the black brane limit at the end of this section.

% --------------------------------------------------------------------------

\subsection{Applying Our Method}

To begin, let us write the partition function of a real scalar in the quasinormal mode representation. From equation (\ref{eq:DHS}), the logarithm of the partition function is
\begin{eqnarray}
\ln Z &=& \Pol + \sum_{\omega_\star}\text{Re}\left[\frac{1}{2}\ln\!\left(\frac{i\omega_\star}{2\pi T}\right) + \ln\!\left(\Gamma\Bigl(\frac{i\omega_\star}{2\pi T}\Bigr)\right)-\frac{1}{2}\ln(2\pi)\right] ,
\end{eqnarray}
where we have incorporated a factor of $1/2$ in order to describe a
real rather than complex scalar.  In addition, we are now denoting the quasinormal frequencies as $\zs=\omega_\star$ and have assumed $\zsbar=\bar\omega_\star = (\omega_\star)^*,$ where an asterisk refers to complex conjugation. This assumption is true for the BTZ scalar quasinormal mode frequencies, which are given by
\begin{equation}
\omega_{k,n,\pm} = \pm \frac{k}{L} - 2\pi Ti(\Delta+2n), \qquad n=0,1,2,\cdots, \,\,\,\,\, k = 0, \pm 1, \pm2, \cdots,
\end{equation}
where $n$ and $k$ are dimensionless numbers which label the mode number the momentum around the spatial circle, respectively.

Using these frequencies, the fixed-$k$ contribution (\ref{eq:logZk})
to the partition function for a real scalar is given by
\begin{align}\label{eq:QNMBTZLogZk}
\ln Z_k&= \Polk + \sum_{n=0}^\infty \left[-\ln(2\pi) + \text{Re}\left(\ln(2n+\Delta+i \hat k)\right) + 2\,\text{Re}\left(\ln\Gamma(2n+\Delta+i \hat k)\right)\right],
\end{align}
where we have defined 
\begin{equation} \label{eq:khat}
\hat k = \frac{k L}{r_h} = \frac{k}{2\pi TL}.
\end{equation}
We will first discuss $\ln Z_k$ and will return to the sum on $k$ later.

% ...........................................................................

\subsubsection{Regularizing the QNM Sum}

In order to regulate the large-$n$ divergence in (\ref{eq:QNMBTZLogZk}) we perform a simple subtraction as in (\ref{eq:Zksub}). In particular, we define a subtracted sum by explicitly removing the terms which diverge as a sum on $n$ at fixed $k$. (The extraction of the divergent terms is especially easy in this
case, since the quasinormal modes are known exactly.  In cases where they
are not, one would need to use WKB for large $n$ to get the necessary
subtractions.) The resulting BTZ expression corresponding to (\ref{eq:ZksubwPol}) is
\begin{align}\label{eq:QNMBTZsubLogZk}
\ln Z_k = \Poltk &+ \ln Z^{\text{sub}}_k \nn\\
= \Poltk &+ \sum_{n=0}^\infty \left[-\ln(2\pi) + \text{Re}\left(\ln(2n+\Delta+i \hat k)\right) + 2\,\text{Re}\left(\ln\Gamma(2n+\Delta+i \hat k)\right)\right] \nn \\
& - \sum_{n=1}^\infty\left[2(2n+\Delta) \ln(2n) - 4n + \frac{1}{12n}\left(1+6(\Delta^2-\hat k^2)\right)\right],
\end{align}
where the second line is determined by taking the large $n$ limit of the summand in (\ref{eq:QNMBTZLogZk}),%
\footnote{
  The two sums in (\ref{eq:QNMBTZsubLogZk}) should be understood as being
  combined into a single (convergent) sum over $n$, with no contribution
  from the second summand for $n=0$.
  Note that,
  since the goal of our subtraction is to cancel the divergence coming from
  large $n$, we could choose the lower limit on $n$ in the second sum of
  (\ref{eq:QNMBTZsubLogZk}) however
  we find convenient.  Choosing a lower limit of $n=2$ instead of $n=1$,
  for example, could be absorbed into a redefinition of $\Poltk$.
  We have avoided choosing a lower limit of $n=0$ because of the
  $1/n$ term in our large-$n$ expansion.  We could have alternatively
  chosen to expand in $1/(n+1)$, again absorbing the difference
  into $\Poltk$.  That would have worked just as well and allowed
  $n=0$ as the lower limit.
}
including all terms up to $O(1/n)$. This sum gives a regularized version of $\ln Z_k$. Note that all of the subtraction terms are explicitly polynomials of $\Delta$. As such, these can be absorbed into $\Polk$ and the difference with (\ref{eq:QNMBTZLogZk}) is absorbed into $\Poltk$.

We now turn to determining $\Poltk$ by matching the large-$\Delta$ asymptotics of (\ref{eq:QNMBTZsubLogZk}) to a regularized calculation of $\ln Z_k$.
First, we need the large $\Delta$ limit of (\ref{eq:QNMBTZsubLogZk}).
Here, large $\Delta$ means $\Delta \gg 1$ and $\Delta \gg \hat k$, but,
because $n$ is summed over, we cannot make any assumption about the size
of $\Delta$ relative to $n$.  Extracting this limit is made easier in
the BTZ case by the fact that we have exact formulas for the frequencies
and so a completely analytic formula for the summand of the first sum in
(\ref{eq:QNMBTZsubLogZk}).  Because of this, we can easily find a
completely analytic result for the large $\Delta$ limit, which is
\begin{eqnarray}
\ln Z_k\Big|_{\Delta\rightarrow\infty}  &= & \Poltk + 2\Delta \ln\Delta -2\Delta \nn \\
&&\quad+ \sum_{n=1}^\infty \left[2(2n+\Delta)\ln\left(1+\frac{\Delta}{2n}\right) - 2\Delta - \frac{\Delta^2}{2n} - \left(\frac{1}{6}-\hat k^2\right)\frac{\Delta}{2n(2n+\Delta)}\right] \nn\\
\label{eq:largeDeltaQNM}
&= & \Poltk + \frac{1}{2}(\hat k^2 - (\Delta-1)^2 + \frac{1}{6})\ln\Delta + \frac{1}{2}(\ln 2 + \frac{3}{2}-\gamma)\Delta^2 \nn \\
&&\quad - (1 +\ln\pi)\Delta + \frac{1}{12}(5-6\hat k^2)\ln2  - \frac{1}{12}\gamma(1-6\hat k^2) - 4\ln A,
\end{eqnarray}
where
$A \equiv \exp\bigl(\ft1{12} - \zeta'(-1)\bigr)$ is the Glaisher constant and $\zeta(x)$ the Riemann $\zeta$-function.

In cases where exact frequencies are not known, we would need to either
(i) get an analytic result for the large $\Delta$ limit of
$\ln Z_k^{\rm sub}$
by devising
a WKB-like analysis of the frequencies that was valid for large $\Delta$
and {\it any}\/ value of $n$ (large, small, and in between), or
(ii) evaluate the analog $\ln Z_k^{\rm sub}$ numerically for
large $\Delta$ and use that to numerically extract the polynomial
$\Poltk$ in the
matching procedure that will follow.
Since our goal here is just to test the structure of our method,
we will just stick with the relatively simple derivation
(\ref{eq:largeDeltaQNM}) for the BTZ case.

In order to determine $\Poltk$, (\ref{eq:largeDeltaQNM}) now needs to be matched to an appropriately regularized calculation of $\ln Z_k$. Following the procedure outlined earlier in section \ref{sec:regfixedk}, we will use a fixed-$k$ heat kernel regularization.

% ...........................................................................

\subsubsection{Fixed-$k$ heat kernel expansion}\label{sec:BTZfixedk}

We need to evaluate the effective two dimensional fixed-$k$ heat kernel (\ref{eq:fixedkHKexp}) which arises from the BTZ background. This expansion will be used {\it both}\/ for (i) finding the large-$\Delta$ limit in order to extract $\Poltk$ and (ii) regulating the large-$k$ asymptotics of the partition function for fixed $\Delta$. Using (\ref{eq:fixedkHKcoeffs1}) and the power counting of section \ref{sec:kHK}, the expansion is formally
\begin{align}
K_k(x;\tk) &= \frac{1}{4\pi\tk}\, e^{-\tk (m^2-E)}\left[1 + \frac{\tk}{6}\left(\Rtw + \tk E_{;\mu}{}^\mu + \frac{1}{2}\tk^2 E_{;\mu} E^{;\mu}\right) + O(\tk_{\rm eff}^2)\right],
\end{align}
which shows all terms we'll need for the $d{=}2$ case of BTZ.
[We've included the subscript ``(2)'' above as a reminder that
the metric and curvature tensors of section \ref{sec:kHK} were
with respect to the two-dimensional geometry of $(r,\tau)$.]

It's useful to reorganize this expansion slightly, first by isolating
the $k^2$ term of E.  Defining $\tilde E$ and $X$ by
$E = \tilde{E} - k^2/r^2 = \tilde{E} - k^2 X$
separates the potentially large $k^2$ term from the rest.  The
expansion can then be rewritten as
\begin{align}\label{eq:BTZfixedkHK0}
K_k(x;\tk) &= \frac{1}{4\pi\tk}\, e^{-\tk(k^2 X + m^2)}\left[1 + \frac{\tk}{6}\left(\Rtw + 6\tilde{E} - \tk k^2 X_{;\mu}{}^\mu + \frac{1}{2}\tk^2 k^4 X_{;\mu}X^{;\mu}\right) + O(\tk_{\rm eff}^2)\right] ,
\end{align}
where
\begin{eqnarray}
	\tilde{E} &=&- \frac{(d-3)(d-1)}{4r^2}f(r) - \frac{2(d-1)}{4r}f'(r), \label{eq:tE} \\
	X&=& \frac{1}{r^2}, \label{eq:X} \\
	X_{;\mu}{}^\mu&=& \frac{6}{r^4}f(r) - \frac{2}{r^3}f'(r),\\
	X_{;\mu}X^{;\mu} &=& \frac{4}{r^6}f(r),\\
	\Rtw&=&  - f''(r) \,.
\end{eqnarray}

The exponential in (\ref{eq:BTZfixedkHK0}) would be awkward if we
happen to be interested in the
case of negative $m^2$ since then, no matter how large $k$ is,
$\exp[-\tk(k^2/r^2+m^2)]$ would be a {\it growing} exponential in $\tk$ for
large enough values of $r$ (i.e.\ close enough to the boundary).
We find it convenient to instead reorganize the expansion in terms of a shifted
mass
\begin {equation} \label{eq:mhat}
\hat m^2 \equiv m^2 + \frac{d^2}{4L^2} = \frac{(\Delta - \tfrac{d}{2})^2}{L^2}.
\end {equation}
Then $\hat m^2$ is positive for all scalar perturbations with $m^2 > m_{BF}^2$, where $m^2_{BF}=-d^2/4L^2$ is the Breitenlohner-Freedman (BF) bound \cite{Breitenlohner:1982jf} for stable scalar perturbations in asymptotically AdS spacetimes. 
A very useful property of the shifted mass $\hat m$, which will simplify matters later on, is that it is analytic (and in particular polynomial) in
$\Delta$, with $\hat m L=(\Delta - d/2)$.
Switching from $m$ to $\hat m$, we rewrite the expansion as
\begin{align}\label{eq:BTZfixedkHK}
K_k(x;\tk) &= \frac{1}{4\pi\tk}\, e^{-\tk(k^2 X + \hat m^2)}\left[1 + \frac{\tk}{6}\left(\Rtw + 6\tilde{E} + \frac{3d^2}{2L^2} - \tk k^2 X_{;\mu}{}^\mu + \frac{1}{2}\tk^2 k^4 X_{;\mu}X^{;\mu}\right) + O(\tk_{\rm eff}^2)\right] ,
\end{align}

In order to compute the logarithm of the partition function we must integrate the heat kernel over $\tk$ as in (\ref{eq:detD}). Integrating (\ref{eq:BTZfixedkHK}) over $\tk$ with a UV cut-off $\tk \gtrsim 1/\Lambda^{2},$ expanding for large $\Lambda$ and dropping terms which vanish as $\Lambda{\to}\infty,$ we have
\begin{align}\label{eq:HKBTZtint}
&
\int_{\Lambda^{-2}}^\infty \frac{d\tk}{\tk} K_k^{\rm trunc}(x;\tk)=
\frac{1}{4\pi}\Biggl\{\Lambda^2 \nn \\
&\qquad
+\bigg(\frac{k^2 + \hat m^2 r^2}{r^2} + \frac{f''(r)}{6} + \frac{d-1}{2r}f'(r)+\frac{(d-1)(d-3)}{4r^2}f(r) - \frac{d^2}{4L^2} \bigg)
\bigg[\ln\left(\frac{k^2+\hat m^2r^2}{\Lambda^2r^2}\right)+\gamma\bigg] \nn \\
&\qquad
- \frac{k^2 + \hat m^2 r^2}{r^2}
- \frac{1}{3r^2}\bigl(3f(r)-r f'(r)\bigr)\frac{k^2}{k^2+\hat m^2r^2} + \frac{f(r)}{3 r^2}\frac{k^4}{(k^2+\hat m^2r^2)^2}\Biggr\}.
\end{align}

Next we must integrate over the two-dimensional spacetime (\ref{eq:2Dslice}). This yields the truncated large-$k$ expansion $F(k)$ of $\ln Z_k$ [defined
by (\ref{eq:IntHK})].  Specializing to the $d{=}2$ case of BTZ with $f(r) = \frac{r^2}{L^2}(1- \frac{r_h^2}{r^2})$, equation (\ref{eq:HKBTZtint}) gives%
\footnote{
  We have split the logarithm up in (\ref{eq:HKBTZintgrtdLogZk})
  just for the convenience of clearly
  separating the UV-divergent $\Lambda$ dependence from the terms
  that depend on $k$.
}
\begin{align}\label{eq:HKBTZintgrtdLogZk}
F(k) \equiv \ln Z_k^{\rm trunc} &= \frac{1}{2}\int_{r_h}^{r_b} dr\int_0^{1/T} d\tau \int_{\Lambda^{-2}}^\infty \frac{d\tk}{\tk}\, K^{\rm trunc}_k(x;\tk) \nn\\
&=\frac{1}{4}\bigg[-\Lambda^2L^2 - \bigg(\hat k^2 - \hat m^2L^2 + \frac{1}{6} \bigg)\ln(\Lambda^2L^2)  - 3\hat k^2 + \hat m^2L^2 \nn\\
& \qquad + \left(\hat k^2 - \hat m^2L^2 + \frac{1}{6} \right)\bigl(\ln(\hat k^2 + \hat m^2L^2) + \gamma\bigr) + 4 \hat k \hat mL \arctan\bigg(\frac{\hat k}{\hat m L}\bigg)\bigg],
\end{align}
where we have used $r_h = 2\pi T L^2$ and again defined $\hat k$
as in (\ref{eq:khat}).

In (\ref{eq:HKBTZintgrtdLogZk}) we have regulated the boundary divergence by cutting off the upper limit of the $r$ integral at some $r_b\gg1$ and then taken the $r_b \rightarrow \infty$ limit while discarding terms proportional to $r_b$.  In particular, we have dropped the divergent boundary term  
\begin {equation}\label{eq:IRdiv}
	\frac{r_b}{8\pi T} \left[\Lambda^2 - \hat m^2 +\left(\hat m^2 + \frac{1}{12L^2}\right)\ln\left(\frac{e^\gamma\hat m^2}{\Lambda^2}\right)\right]
\end {equation}
from (\ref{eq:HKBTZintgrtdLogZk}).
If one prefers, one may get the same result (i.e.\ dropping the power
law divergence in $r_b$) by using dimensional regularization in the
gravity theory.  (Note that this would correspond to using dimensional
regularization for the IR behavior of the gravity theory, while we
are using the more common heat kernel regularization with $\Lambda$
to cut off the
UV behavior of the gravity theory.  There's no reason one can't use
both.)
The proof, perhaps, is in the pudding: We will see that
this prescription for boundary regularization indeed gives the correct
result for the partition function.

The result (\ref{eq:HKBTZintgrtdLogZk}) will prove useful in both determining $\Poltk$ and in regulating the large-$k$ asymptotics of the partition function. For now, we focus on the former use and take the $\Delta{\to}\infty$ limit to obtain
\begin{align}\label{eq:largeDeltaHK}
(\ln Z_k)\Big|_{\Delta\rightarrow\infty} & =  \frac{1}{4}\bigg[-\Lambda^2L^2 -  \left(\hat k^2 - (\Delta-1)^2 + \frac{1}{6} \right)\ln(\Lambda^2L^2) \nn \\
& \qquad + \left(\hat k^2 - (\Delta-1)^2 + \frac{1}{6} \right)\left(2\ln \Delta + \gamma\right) + \Delta^2 - 2\bigg].
\end{align}
We determine $\Poltk$ by comparing this with the large $\Delta$ limit of the DHS QNM sum in (\ref{eq:largeDeltaQNM}). This gives
\begin{eqnarray}\label{eq:Poltk}
\Poltk &=& -\frac{1}{4}\Lambda^2L^2 - \frac{1}{4}\left(\hat k^2 - (\Delta-1)^2 + \frac{1}{6} \right)\ln(\Lambda^2L^2) \nn \\
&&  - \frac{\Delta^2}{4}(2\ln2 + 2 - \gamma) + \frac{\Delta}{2}(2\ln\pi+2+\gamma) + \frac{\hat k^2}{4}(2\ln2-\gamma) \nn\\
&& -\frac{1}{2} - \frac{\gamma}{8} - \frac{5}{12}\ln2 + 4\ln A.
\end{eqnarray}
Note that the large $\Delta$ expressions (\ref{eq:largeDeltaQNM}) and (\ref{eq:largeDeltaHK}) both contain terms which are not meromorphic in $\Delta.$ In particular, they have $\ln \Delta$ dependence. One key assumption in the formalism of \cite{Denef:2009kn} is that the quasinormal modes determine the log of the partition function up to a local {\it polynomial} in $\Delta.$ Similar to the case of even dimensional de-Sitter spaces discussed in \cite{Denef:2009kn,Keeler:2014hba}, the fact that the non-meromorphic $\Delta$ dependence in the heat kernel and quasinormal mode representation of the partition function cancel when computing $\Poltk$ provides a non-trivial consistency check of the application of such techniques.  In our case, it provides a non-trivial consistency check to the application to the fixed-$k$ partition function. In particular we see that our $\Poltk$ is a polynomial in $\Delta$, as required.

Plugging this expression for $\Poltk$ into (\ref{eq:QNMBTZsubLogZk}) and summing over momentum modes gives the complete partition function. However, as discussed earlier, the sum over momenta is divergent. To regularize this divergence we add and subtract the truncation of the fixed-$k$ heat kernel expansion (\ref{eq:BTZfixedkHK}), as described in section \ref{subsec:Detdet}. In this example, the function $F(k)$ in (\ref{eq:logZfinal}) is given by (\ref{eq:HKBTZintgrtdLogZk}).

Expression (\ref{eq:HKBTZintgrtdLogZk}) implicitly assumes that $k$ is small
compared to the UV momentum cutoff determined by $\Lambda$.
As discussed back in section \ref{subsec:Detdet}, this assumption is
adequate except for the computation of the integral $I = \int_k F(k)$.
Our UV regularization $\Lambda$ was introduced in integration over
the heat kernel parameter $\tk$, as in (\ref{eq:HKBTZtint}).
We find that the simplest way to allow for $k$ of order the
UV momentum cutoff is to go back
and sum over $k$
{\it before}\/ the integral over $\tk$ when computing $I$.
In Appendix \ref{app:PoissonResum}, we show how to employ Poisson resummation to compute $I$ for general $X$. For our specific case (\ref{eq:X}) of $X = 1/r^2$, we find the simple result
\begin{eqnarray}\label{eq:HKsummed}
I &=&\int d^3x \sqrt{g}\left(\frac{\Lambda^3}{24\pi^{3/2}} -\frac{(\Delta-1)^2\Lambda}{8\pi^{3/2}L^2} + \frac{(\Delta-1)^3}{12\pi L^3}\right) \nn \\
&& + \frac{1}{(2\pi)^2}\frac{1}{(2\pi TL)^2}\, \Li_3 (e^{-4\pi^2 (\Delta-1)TL}) - \frac{1}{12}\, \Li_1(e^{-4\pi^2 (\Delta-1)TL}),
\end{eqnarray}
where $g$ is the metric determinant of the three-dimensional spacetime, $\Lambda$ is a UV cutoff introduced in the same way as in (\ref{eq:HKBTZtint}), and $\Li_n(x)$ are poly-logarithms which are defined by
\begin{equation}
\Li_n(x) = \sum_{k=1}^\infty \frac{x^k}{k^n}.
\end{equation}

In the first line of (\ref{eq:HKsummed}) we have recovered the usual asymptotics of the partition function in standard heat kernel regularization. This agrees with the local terms given in \cite{Denef:2009kn}. The second line in (\ref{eq:HKsummed}) however, contains finite contributions to the partition function. These are non-zero at finite temperature and must be included in order to match to previous results on the BTZ scalar partition function.
We can now put together the various contributions to the partition function.

% ----------------------------------------------------------------------------

\subsection{A brief aside on an earlier warning}
\label {sec:warning}

Before we put everything together,
we should explain a subtlety of our formula for
$F(k)$.  In the large-$k$ limit (for fixed $\hat m$), the last term
$\frac14 \times 4 \hat k \hat mL \arctan(\hat k/\hat m L)$
of (\ref{eq:HKBTZintgrtdLogZk}) becomes
\begin {equation}
  \tfrac{\pi}{2} |\hat k| \hat m L .
\label {eq:questionable}
\end {equation}
This looks a little different than the other
terms in (\ref{eq:HKBTZintgrtdLogZk}) because it depends on
$\hat m$ instead of $\hat m^2$.
In fact, we show in appendix \ref{app:largishr} that this particular
term is generated by the region of the
$r$ integral in (\ref{eq:HKBTZintgrtdLogZk}) for which
$r \sim k^2/\hat m^2$, which is large when $k$ is large.
This $r$ is large enough that the large-$k$ expansion derived in
section \ref{sec:kHK} cannot be trusted (for non-large $\hat m$),
as we warned earlier.
The other terms in (\ref{eq:HKBTZintgrtdLogZk}), in contrast,
turn out to come from $r \sim r_h$, for which all is well.

So what to do?  Note that (\ref{eq:questionable}) is polynomial in
$\Delta$ because $\hat m L=(\Delta - d/2)$ is.  So, if we wanted,
we could simply redefine $F(k)$ to drop the troublesome
term (\ref{eq:questionable}) altogether and then
exactly absorb that change into a corresponding redefinition of the
polynomial $\Poltk$.  However we move things around between
$F(k)$ and $\Poltk$---whether we keep the troublesome term
in $F(k)$ or drop it---we will
get the same result for the combination (\ref{eq:logZfinal}).
This suggests that it may not really matter whether we get
the particular term (\ref{eq:questionable}) wrong, as long as
it's a polynomial in $\Delta$.
And that's the advantage to using
the shifted mass $\hat m$ in the calculation instead of $m$,
since the latter is not polynomial in $\Delta$.

We will indeed see that the above suggestion is born out:
In the next subsection, we verify that
blindly using (\ref{eq:HKBTZintgrtdLogZk}) for $F(k)$
correctly reproduces the known BTZ partition function.
However, we would like an argument other than
answer-analysis that
this procedure should work, so that we know it is not a special
property of the BTZ black hole.  Our problem occurs at large $r$,
where the space-time is well approximated by AdS.
In appendix \ref{app:largishr}, we show that
(\ref{eq:questionable}) actually {\it does} corresponds to the exact answer
for $\log Z_k$ in locally AdS$_3$ space-time.

% ----------------------------------------------------------------------------

\subsection{Final form of the partition function}

Inserting the expressions derived in this section into equation (\ref{eq:logZfinal}) gives the final result for the logarithm of the partition function,
\begin{eqnarray}\label{eq:finalLogZ}
\ln Z &=& \int d^3x \sqrt{g}\left(\frac{\Lambda^3}{24\pi^{3/2}} -\frac{(\Delta-1)^2\Lambda}{8\pi^{3/2}L^2} + \frac{(\Delta-1)^3}{12\pi L^3}\right) \nn \\
&& + \frac{1}{(2\pi)^2}\frac{1}{(2\pi TL)^2}\, \Li_3 (e^{-4\pi^2 (\Delta-1)TL}) - \frac{1}{12}\, \Li_1(e^{-4\pi^2 (\Delta-1)TL}) \nn \\
&& + \sum^{\infty}_{k=-\infty} \left( \Poltk + \ln Z_k^{\text{sub}} - F(k)\right),
\end{eqnarray}
where $\ln Z_k^{\text{sub}},$ $F(k)$ and $\Poltk$ are given in  (\ref{eq:QNMBTZsubLogZk}), (\ref{eq:HKBTZintgrtdLogZk}) and (\ref{eq:Poltk}), respectively. Note also that all UV-divergent terms are included on the first line above since the $\Lambda$-dependent divergences explicit in $F(k)$ and $\Poltk$ exactly cancel [as can be seen by comparing equations (\ref{eq:HKBTZintgrtdLogZk}) and (\ref{eq:Poltk})].

In order to compute (\ref{eq:finalLogZ}) there are two sums to perform. In particular, in addition to the explicit sum on $k$, recall that $\ln Z_k^{\text{sub}}$ contains a sum over mode numbers labeled by $n$.  We do not know how to perform
these sums analytically, but remember that our motivation was to propose a method that could be used numerically for other black hole spacetimes.  The BTZ calculation here is offered simply as a check.  We move now to demonstrating that computing (\ref{eq:finalLogZ}) numerically indeed recovers the expected result for $\ln Z$ by comparing to the results of \cite{Denef:2009kn}.

% ----------------------------------------------------------------------------

\subsection{Comparison to DHS \cite{Denef:2009kn}}

The scalar partition function in the BTZ background has been previously computed in \cite{Denef:2009kn,Mann:1996ze}. The results of \cite{Denef:2009kn} are particularly straightforward for comparison as they derive the partition function using the same heat kernel regularization as we have above. The result of \cite{Denef:2009kn} is%
\footnote{
Notice that in equation (63) of \cite{Denef:2009kn}, DHS have absorbed the $\Lambda^3$ term into the overall constant contribution to $\ln Z$.  This is the overall normalization of the partition function, which is undetermined by the heat kernel. In contrast, in (\ref{eq:finalLogZ}) we have implicitly set to zero the corresponding ``$\text{const.}$'' introduced just before DHS (63) and have explicitly kept the leading $\Lambda^3$ divergence in $\ln Z$.
}
\begin{eqnarray}\label{eq:DHSBTZlogZ}
\ln Z &=& \text{const.} + \int d^3x \sqrt{g}\left(\frac{\Lambda^3}{24\pi^{3/2}}
 -\frac{(\Delta-1)^2\Lambda}{8\pi^{3/2}L^2} + \frac{(\Delta-1)^3}{12\pi L^3}\right) \nn\\
&&+ \ln\prod_{\kappa=0}^{\infty}(1-q^{\kappa+\Delta})^{-(\kappa+1)},
\end{eqnarray}
where $q = e^{-4\pi^2 TL}$.
The first lines of (\ref{eq:finalLogZ}) and (\ref{eq:DHSBTZlogZ}).  So, in order to check our representation of the partition function, we should compare the last two lines of (\ref{eq:finalLogZ}) with the second line of (\ref{eq:DHSBTZlogZ}).

\begin{figure}
	\centering
	\includegraphics[width=.7\textwidth]{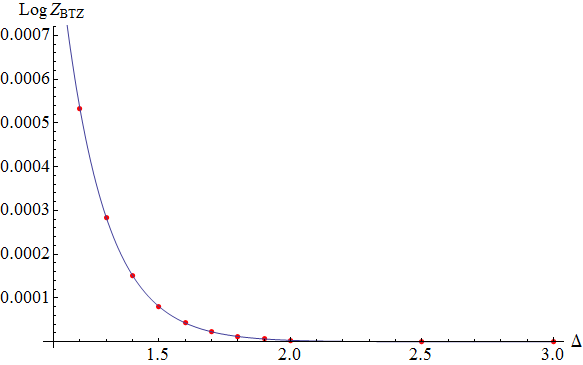}
	\caption{Plot of $\ln Z(\Delta).$ The red dots are the numerical results of the last two lines of (\ref{eq:finalLogZ}) for $\Delta = \{1.1,1.2,1.3,\cdots,1.9,2.0,2.5,3.0\}$ and with $2\pi TL = 1.$ The blue line is a plot of the logarithm of the finite temperature partition function in the second line of (\ref{eq:DHSBTZlogZ}) at the same value of $TL.$}\label{fig:LogZplot}
\end{figure}

\begin{figure}
	\centering
	\includegraphics[width=.7\textwidth]{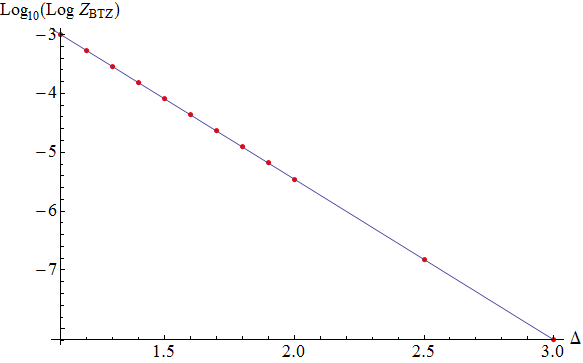}
	\caption{Plot of $\log_{10}(\ln Z(\Delta)).$ The red dots are the numerical results from the last two lines of (\ref{eq:finalLogZ}) for $\Delta = \{1.1,1.2,1.3,\cdots,1.9,2.0,2.5,3.0\}$ and with $2\pi TL = 1.$ The blue line is a plot of $\log_{10}$ of the logarithm of the finite temperature partition function in the second line of (\ref{eq:DHSBTZlogZ}) at the same value of $TL.$}\label{fig:LogLogZplot}
\end{figure}

We compute both sums numerically.\footnote{In practice, when numerically computing the sum in (\ref{eq:finalLogZ}) we included extra subtraction terms in $\ln Z^{\text{sub}}_k$ than are explicitly shown in (\ref{eq:QNMBTZsubLogZk}). In particular, we subtract terms corresponding to higher order powers of $1/n$ in the expansion of the summand in (\ref{eq:QNMBTZLogZk}) that are convergent as a sum on $n$ in order to improve the rate convergence of the numerical sum. Since these terms have convergent sums we simply add back the analytic result for them by hand.} The results are illustrated in Fig. \ref{fig:LogZplot} and Fig. \ref{fig:LogLogZplot}. In Fig. \ref{fig:LogZplot} we plot $\ln Z$ directly and compare to the results of \cite{Denef:2009kn}. Since the result approaches zero rapidly as $\Delta$ increases, we present the corresponding log plot in Fig.\ \ref{fig:LogLogZplot}, which clearly shows agreement up to $\Delta=3.$

Finally, one can perform precisely the same calculation for the case where the horizon is an infinite spatial line instead of a circle. As mentioned previously, this corresponds to the same manipulations as above, except that the sum over momentum modes is replaced by an integral. The analytic results for the integrals are worked out in Appendix \ref{app:HK}. As is straightforward from the results of Appendix \ref{app:HK}, the end result is the same as equation (\ref{eq:finalLogZ}), except that the finite terms in the second line of (\ref{eq:finalLogZ}) are absent. Numerically evaluating the integral of the last line of (\ref{eq:finalLogZ}) (instead of the sum) we find that the integral vanishes to within the accuracy we computed. This is consistent with the $TL\rightarrow \infty$ limit of the last line of (\ref{eq:DHSBTZlogZ}) and with the expectations that, in the de-compactification limit, the finite temperature contributions to the free energy should vanish. This fact can also be seen by taking the large temperature limit $TL \to \infty$ in (\ref{eq:DHSBTZlogZ}).

% ============================================================================

\section{Discussion}\label{sec:disc}

In this paper, we have presented a procedure to compute numerically the partition function of fluctuations about asymptotically anti-de Sitter black holes using the quasinormal mode spectrum. We illustrated the method by computing the scalar partition function in the BTZ black hole and reproduced the known result. Our method provides a straightforward generalization of the method proposed in \cite{Denef:2009kn} to cases in which the quasinormal mode spectrum is not known analytically.  The key new ingredient is the development of the fixed momentum partition function and corresponding heat kernel.

There are many obvious extensions of this current work. First, we have only considered scalars in the BTZ black hole. It is natural to consider other spin fields and develop the corresponding fixed-$k$ heat kernel, which should be straightforward.

A more ambitious goal is to apply this methodology to higher dimensional asymptotically AdS black holes. This is the main motivation for our work and would provide a non-trivial test and application of our proposed method to a scenario in which the quasinormal modes are not known analytically. There are several potential difficulties in performing such a calculation. First, one has to determine the asymptotic values of the quasinormal frequencies as in (\ref{eq:WKBQNM}). While this can be done in a WKB approximation, the calculation requires going to higher subleading orders in the inverse mode number $1/n$ than have so far been computed in the literature, in order to ensure convergence of the sum in $\ln Z_k^{\text{sub}}.$ Second, having such a result, one needs to determine $\Polk.$ If this can only be done numerically then, in order to reliably fit to a numerical result of $\Polk$, it would be beneficial to have an understanding of the expected dependence of this function on $k.$ Based on our experience with BTZ, it appears likely that (at least for scalar fields) $\Polk$ is a polynomial in both $\Delta$ and $k^2$. However, it would be desirable to have an analytic argument for such functional dependence of $\Polk$ on $k^2.$ Another hope is that an appropriate WKB expression of the quasinormal modes can be determined in the limit of large-$\Delta$ which is valid for arbitrary mode number $n.$ This would interpolate between the large-$\Delta$ result of \cite{Festuccia:2008zx} for small values of $n$ and the large-$n$ results of \cite{Natario:2004jd} for asymptotically large values of $n.$ Armed with such an expression one should be able to determine $\Polk$ analytically, and we are currently investigating this possibility. Clearly, much work is necessary to extend the current results to more interesting examples, and we hope to turn to such calculations in the near future.

%============================================================================
%============================================================================

\acknowledgments

DV would like to thank the Dept. of Physics, College of William and Mary, where part of this work has been completed, for hospitality. PS would like to thank G.~Festuccia and C.~Keeler for useful correspondence. This work is part of the D-ITP consortium, a program of the Netherlands Organisation for Scientific  Research (NWO) that is funded by the Dutch Ministry of Education, Culture and Science (OCW), and is also supported in part by the US Department of Energy under grant DE-SC0007894.

%============================================================================
%============================================================================
\appendix
%============================================================================

\section{Derivation of fixed-\boldmath$k$ heat kernel coefficients}\label{app:HK}

In this appendix we provide the details of the derivation of the fixed-$k$ heat kernel expansion. The heat kernel $K(x,x';\tk)$ satisfies the heat equation
\begin{equation}\label{eq:heat}
(\partial_\tk + D_x)\, K(x,x';\tk) = 0,
\end{equation}
where $D_x = -(\nabla^2 +E).$ We will be particularly interested in the case where $E$ depends on a parameter that can become parametrically large. In our application in the main text, $E$ has a term proportional to $k^2,$ where $k$ can be thought of as momentum eigenvalues for mode functions along the space transverse to the $r$-$\tau$ plane. Small values of $\tk$ in the heat kernel correspond to high energies. When $k^2$ becomes parametrically large and of the order of $1/\tk$ as $\tk\rightarrow0$ we will need to solve (\ref{eq:heat}) in an expansion that remains valid for such large values of momentum. For our purposes this means we will need a solution $K_k(x,x';\tk)$ as an expansion for small $\tk$ while allowing $\tk k^2 \sim O(1)$ or, equivalently, $\tk E \sim O(1)$. 

To illustrate how the potential $E$ affects the heat kernel expansion, first consider the usual case where $E$ is independent of $k.$ There is an elegant solution to the heat equation due to DeWitt \cite{DeWitt:1965jb} which in $(d{+}1)$-dimensions takes the form
\begin{equation}\label{eq:deWitt}
K(x,x';\tk) = (4\pi\tk)^{-(d+1)/2} \DD(x,x')\,e^{-\frac{\sigma(x,x')}{2t}}\,\Xi(x,x';\tk).
\end{equation}
In (\ref{eq:deWitt}), $\sigma(x,x')$ is one-half of the square of the geodesic distance between $x$ and $x'$ 
\begin{equation}\label{eq:sigma}
\sigma(x,x') = \frac{1}{2}\left( \int_{x'}^{x}\sqrt{g_{\mu\nu}(\bar x)d\bar x^\mu d\bar x^\nu}\right)^2,
\end{equation}
where the path of integration is given by the geodesic connecting $x$ to $x'.$ Alternatively, this can also be written in terms of Synge's world function
\begin{equation}
\sigma(x,x') = \frac{1}{2}(\lambda_1-\lambda_0) \int_{\lambda_0}^{\lambda_1}g_{\mu\nu}(\bar x(\lambda)) \, t^\mu t^\nu \, d\lambda,
\end{equation}
where $\bar x(\lambda)$ is the geodesic connecting $x=\bar x(\lambda_1)$ and $x'=\bar x(\lambda_0),$ $t^\mu = d\bar x^\mu/d\lambda$ is a tangent vector to the geodesic and $\lambda$ is an affine parameter. In addition, $\Delta(x,x')$ is the van Vleck determinant, which we define as
\begin{equation}
\Delta(x,x') = - \frac{1}{\sqrt{g\,g'}}\det \left[\frac{\partial}{\partial x^\alpha}\frac{\partial}{\partial x'^\beta}\sigma(x,x')\right] \equiv  - \frac{1}{\sqrt{g\,g'}} \det [\sigma_{\alpha\beta'}(x,x')].
\end{equation}
See \cite{Poisson:2011nh} for a detailed discussion of the properties of these and other bi-scalar quantities encountered in the expansion (\ref{eq:deWitt}).

Going back to (\ref{eq:deWitt}), the function $\Xi$ is then expanded in a power series in $\tk$
\begin{equation}\label{eq:HKnormal}
\Xi(x,x';\tk) = \sum_{k=0}^\infty a_{2k}(x,x')\,\tk^k,
\end{equation}
where the bi-scalars $a_{2k}(x,x')$ are called heat kernel coefficients. The coefficients $a_{2k}(x,x')$ can be solved iteratively by inserting the ansatz (\ref{eq:deWitt}) into the heat equation (\ref{eq:heat}). Usually, the potential $E$ is a local function of the coordinates. As long as this function is well behaved, it will not interfere with the expansion in $\tk$. In fact, the leading dependence on $E$ can naturally be seen by considering the heat kernel expansion as an expansion of $\text{Tr}\, e^{-\tk(D_x+m^2)} = \text{Tr}\, e^{-\tk(-\nabla^2 -E +m^2)}$ for small $\tk$. One can choose to factor out the $e^{\tk E}$ from this trace expression.\footnote{This behavior is apparent in the heat kernel coefficients (\ref{eq:HKacoeff}), where one can see that leading $E$ dependent terms (which do not include derivatives of $E$) appear to exponentiate into $e^{\tk E}.$} Note however that this operation does not commute with the trace, and there remains dependence on derivatives of $E$ that is not captured in the $e^{\tk E}$ term. A proper understanding of these derivative terms is crucial in developing the fixed-$k$ heat kernel that we discuss next. 

Now consider a situation, as in the main text, where $E$ has dependence on a parameter which can become parametrically large compared to $\tk^{-1}.$ In the following, we will call this parameter $k$ and assume that it appears quadratically in $E.$ Furthermore, we will assume that this dependence arises from decomposing the space of eigenfunctions of our operator into eigenmodes of a $(d - \df)$-dimensional transverse space such that (after dimensionally reducing on this transverse space) the problem can be formulated in a space of effective dimension $\df{+}1.$ In the application in the main text we will be interested in the case $\df = 1$, but for sake of generality we leave it arbitrary in this appendix. We will refer to the heat kernel in the dimensionally reduced problem as the fixed-$k$ heat kernel and denote it by $K_k(x,x';\tk),$ where here and in what follows all bi-scalar quantities will be defined with respect to the $(\df{+}1)$-dimensional geometry. 

Consider the modified deWitt ansatz
\begin{equation}\label{eq:deWittmod}
K_k(x,x';\tk) = (4\pi\tk)^{-(\df+1)/2} \DD(x,x') \, e^{-\frac{\sigma(x,x')}{2\tk} + \tk E(x)}\,\Theta(x,x';\tk),
\end{equation}
where $x$ and $x'$ are now coordinates on the effective $(\df+1)$-dimensional space. We have chosen to keep the $\tk E$ term in the exponential. This keeps an explicit term schematically of the form $e^{-\tk k^2}$ in the heat kernel. This term both removes the leading $k^2$ dependence of the heat kernel coefficients and effectively provides a cut-off for the large-$k$ modes. We now seek a series solution for $\Theta(x,x';\tk)$ by expanding as a power series in $\tk$ as
\begin{equation}\label{eq:HKexpmod}
\Theta(x,x';\tk) = \sum_{j=0}^\infty b_{2j}(x,x') \, \tk^j.
\end{equation}
Inserting this ansatz into the heat equation now yields a modified recursion relation for the coefficients $b_{2j}.$ Doing this we find the following recursion relation:
\begin{eqnarray}\label{eq:recursion}
0&=& (j + \sigma_{;\mu} \nabla^\mu)\,b_{2j}(x,x') \nn \\
&& - (\Delta^{-\frac{1}{2}} \nabla^\mu \nabla_\mu \DD - \sigma_{;\mu} E^{;\mu})\,b_{2j-2}(x,x') \nn \\
&& - (2 \Delta^{-\frac{1}{2}} \DD_{;\mu} E^{;\mu} + E_{;\mu}^{;\mu} + 2 E^{;\mu} \nabla_\mu)\,b_{2j-4}(x,x')\nn\\
&& - E_{;\mu}E^{;\mu}\,b_{2j-6}(x,x'),
\end{eqnarray}
where all quantities are defined with respect to the dimensionally reduced space and all explicit covariant derivatives act on everything to their right. When applying the recursion relation it is important to remember that all quantities (except for the potential $E$) are bi-scalars having both $x$ and $x'$ as arguments. We are eventually interested in the coincidence limit of these quantities, which corresponds to the limit $x' \rightarrow x$. However, at intermediate steps it is important to keep the full bi-scalar dependence. Finally, note that setting derivatives of $E$ to zero in (\ref{eq:recursion}) and replacing the $\nabla^2$ term with $\nabla^2 +E,$ one recovers the standard heat kernel coefficient recursion relations (see, for example \cite{Vassilevich:2003xt}).

The benefit of the recursion relation (\ref{eq:recursion}) is that it explicitly contains the $E_{;\mu}$ terms which, at any given order in the small-$\tk$ expansion, can multiply a factor of $\tk$ to give a parametric dependence such that $\tk E \sim O(1)$. This can effectively reduce the order of any given term in the small-$\tk$ expansion as we will see in the following.

% ----------------------------------------------------------------------------

\subsection{Evaluating the heat kernel coefficients}

Before discussing the validity of the above expansion, we will evaluate the first several heat kernel coefficients in order to set up notation for the upcoming discussion and to illustrate the methodology used in solving the recursion relation (\ref{eq:recursion}).

We start the recursion relation by setting $b_{2j}(x,x') = 0$ for $j<0,$ and use the initial condition $b_0(x,x')=1.$ Next, the coincidence limit of $b_2(x,x')$ can be determined directly from (\ref{eq:recursion}) yielding
\begin{equation}\label{eq:b2}
b_2(x,x) = \frac{1}{6} \Rf,
\end{equation}
where we have taken the coincidence limit $x' \rightarrow x$ and we have put a subscript ``(eff)'' to emphasize when tensors are defined in the dimensionally reduced geometry. We will often denote this limit by putting a quantity inside of square brackets.  For example we may write $[b_2] = b_2(x,x) = b_2(x)$. Also, here we used the rudimentary results on coincidence limits for $\sigma$ and $\DD$
that
\begin{align}\label{eq:coin}
[\sigma] = [\sigma_{;\mu}] = [\DD_{; \mu}] &= 0, \nn \\
[\sigma_{;\mu\nu}] &= \gf_{\mu\nu}, \nn \\
[\DD] &= 1, \nn\\
[\DD_{; \mu\nu}] &= \frac{1}{6} \Rf_{\mu\nu}.
\end{align}
These expressions follow from the coincidence limits of derivatives of the defining relations
\begin{eqnarray}\label{eq:sigmaDD}
\sigma_{;\mu}\sigma^{;\mu} &=& 2\sigma, \nn \\
\DD \sigma_{;\mu}{}^{\mu} + 2\sigma^{;\mu} \DD_{;\mu} &=& (\df+1)\DD.
\end{eqnarray}
One can derive relations similar to (\ref{eq:coin}) for the coincidence limit of higher derivatives of $\sigma$ and $\DD$ by further differentiating (\ref{eq:sigmaDD}). The resulting expressions become quite cumbersome, and we do not include them here but refer the reader to \cite{Poisson:2011nh} for further discussion. (See also \cite{Barvinsky:1985an} for similarly useful expressions involving derivatives of the van Vleck determinant.) Finally, before continuing, notice that (\ref{eq:b2}) is not the same as the usual $a_2(x).$ In particular, it is missing a term linear in $E.$ This dependence has instead been included in the exponential $e^{\tk E}$ in (\ref{eq:deWittmod}).

For $b_4(x),$ (\ref{eq:recursion}) gives
\begin{equation}\label{eq:b4}
2 [b_4] = \frac{1}{6}\Rf [b_2] + [b_2{}_{;\mu}{}^\mu] + E_{;\mu}^{;\mu} .
\end{equation}
In order to determine $b_4$ we see that we not only need $b_2$ but we also need its derivatives. The relevant derivatives on $b_2$ have the form
\begin{align}
[b_2{}_{;\mu}{}^\mu] &= -\frac{1}{3}[(\DD)^{;\nu} {}_\nu]^2 +\frac{1}{3}[(\DD)^{;\nu} {}_\nu {}^\mu {}_\mu] -  \frac{2}{3} E^{;\mu}_{;\mu}.
\end{align}
Inserting this and (\ref{eq:b2}) into the expression for $[b_4]$ gives the usual contribution to the heat kernel, where the term proportional to $E^{;\mu}_{;\mu}$ will give an $O(k^2)$ contribution. Explicitly, evaluating (\ref{eq:b4}) we find
\begin{equation}
\lbrack b_4\rbrack = \frac{1}{72}\Rf^2 - \frac{1}{180}\Rf_{\mu\nu}\Rf^{\mu\nu} + \frac{1}{180} \Rf_{\mu\nu\rho\sigma}\Rf^{\mu\nu\rho\sigma} + \frac{1}{30}\nabla^2 \Rf + \frac{1}{6}E_{;\mu}{}^\mu.
\end{equation}
The remaining heat kernel coefficients can be evaluated similarly, and we include their results later in this appendix.

% ----------------------------------------------------------------------------

\subsection{Determining the relevant terms in the heat kernel expansion}

In the following we will focus on the case $d\leq 4.$ In order to incorporate all of the large-$k$ divergences of the partition function we need to keep all terms of at most $O(\tk_{\text{eff}}^2)$ in the heat kernel expansion, where the power of $\tk_\text{eff}$ is determined by the power of $\tk$ in the expansion (\ref{eq:HKexpmod}) after setting $\tk k^2 \sim \tk E \sim O(1).$ As mentioned in the main text, one may worry that this expansion could be contaminated by ever higher powers of $k$ in the heat kernel coefficients. For example, at $O(\tk^n)$ one would naively expect terms of the form
\begin{equation}
b_{2n} \sim \nabla^{2n} e^{\tk E} \sim (\tk k^2)^{2n} e^{\tk E}\sim O(\tk_{\text{eff}}^0),
\end{equation}
which would lead to divergences occurring at all orders in the heat kernel expansion. Factoring out $e^{\tk E}$ as in (\ref{eq:deWittmod}) helps this situation but does not completely remove all large-$k$ terms.

Thankfully, a well behaved expansion does exist. Let us start by analyzing the relevant power counting. From the previous analysis we know that $[b_0]$ and $[b_2]$ are both independent of $k^2$ and $[b_4]$ is proportional to $k^2$. Next, consider the recursion relation (\ref{eq:recursion}). Recalling that in the coincidence limit $[\sigma],$ $[\sigma_{;\mu}]$ and $[\DD_{;\mu}]$ all vanish, one can see that (once the coincidence limit is taken) the leading dependence on $k^2$ of the coefficient $[b_{2j}]$ will be the same as the leading $k$-dependence of $k^2[b_{2j-4}]$ and $k^4 [b_{2j-6}].$ This implies the term in the heat kernel expansion at $j=3$ will scale as $\tk^3[b_6] \sim O(\tk_{\text{eff}}).$

Given the scaling of the leading heat kernel coefficients, and applying the recursion relation argument of the previous paragraph, we see that the terms in (\ref{eq:HKexpmod}) including $\{[b_{6j-4}],[b_{6j-2}],[b_{6j}]\}$ will have leading behavior that scales as $O(\tk_{\text{eff}}^j).$ So, in general, in order to keep terms up to $O(\tk_{\text{eff}}^j)$ one needs to compute up to the leading $k$-dependence of $[b_{6j}],$ which will be proportional to $k^{4j}.$ To include all large-$k$ divergences for $d$ up to $d{=}4$ we should keep up to $O(\tk_{\text{eff}}^2)$ in the expansion. So we need to compute (the relevant $k$-dependence of) all heat kernel coefficients up to the $\tk^6$ coefficient $[b_{12}].$

For completeness, we list below the coincidence limit of the recursion relation (\ref{eq:recursion}) up to the $b_{12}$ term, including only those terms required in order to keep all large-$k$ divergences for $d\leq4.$ Similar to the notation in section \ref{sec:kHK} we will use brackets with a subscript $\partial E^n$ to denote that we keep only terms containing $n$ or more powers of $E$ when taking the coincidence limit for that particular term.  The recursion relations
are
\begin{align}\label{eq:HKcoeffCL}
0&= 2[b_4] - [b_{2;\mu}{}^\mu] - \frac{1}{6}\Rf [b_2] - E_{;\mu}{}^{;\mu}, \nn \\
0&= 3[b_6]_{\dE} - [b_{4;\mu}{}^\mu]_{\dE}- \frac{1}{6}\Rf [b_4]_{\dE}- E_{;\mu}{}^{;\mu}[b_2] - 2 E^{;\mu} [b_{2;\mu}] - E^{;\mu}E_{;\mu}, \nn \\
0&= 4[b_8]_{\dEtw} - [b_{6;\mu}{}^\mu]_{\dEtw}- \frac{1}{6}\Rf [b_6]_{\dEtw}- E_{;\mu}{}^{;\mu}[b_4]_{\dE} - 2 E^{;\mu} [b_{4;\mu}]_{\dE} - E^{;\mu}E_{;\mu}[b_2], \nn \\
0&= 5[b_{10}]_{\dEth} - [b_{8;\mu}{}^\mu]_{\dEth}- \frac{1}{6}\Rf [b_8]_{\dEth}- E_{;\mu}{}^{;\mu}[b_6]_{\dEtw} - 2 E^{;\mu} [b_{6;\mu}]_{\dEtw} - E^{;\mu}E_{;\mu}[b_4]_{\dE}, \nn \\
0&= 6[b_{12}]_{\dEfo} - [b_{10;\mu}{}^\mu]_{\dEfo}- \frac{1}{6}\Rf [b_{10}]_{\dEfo}- E_{;\mu}{}^{;\mu}[b_8]_{\dEth} - 2 E^{;\mu} [b_{8;\mu}]_{\dEth} - E^{;\mu}E_{;\mu}[b_6]_{\dEtw}. 
\end{align}
%
% ----------------------------------------------------------------------------

\subsection{$[\nabla^n b_m]$ relations}
In this subsection we list the results for the relevant heat kernel coefficients and their derivatives required to evaluate (\ref{eq:HKcoeffCL}). When inserted into (\ref{eq:HKcoeffCL}) these reproduce the results (\ref{eq:fixedkHKcoeffs1}) and (\ref{eq:fixedkHKcoeffs2}) quoted in the main text.
\begin{align}
[b_2] &= \frac{1}{6}\Rf \nn \\
[b_{2;\mu}] &= \frac{1}{12}\Rf_{;\mu} - \frac{1}{2}E_{;\mu} \nn \\
[b_{2;\mu\nu}]_{\dE} &= - \frac{2}{3}E_{;\mu\nu} \nn \\
[b_{2;\mu}{}^{\mu}] &= \frac{1}{15} \nabla^2 \Rf - \frac{1}{90} \Rf_{\mu\nu}\Rf^{\mu\nu} + \frac{1}{90} \Rf_{\mu\nu\rho\sigma}\Rf^{\mu\nu\rho\sigma} - \frac{2}{3}E_{;\mu}{}^\mu \nn\\
[b_{2;\mu\nu\rho}]_{\dE} &=  - \frac{3}{4}E_{;\mu\nu\rho} + \frac{1}{6}\Rf_{\rho\mu\nu}{}^\lambda E_{;\lambda} + \frac{1}{12}\Rf_{\mu\nu\rho}{}^\lambda E_{;\lambda} \nn \\
[b_{2;\mu}{}^\mu{}_\nu{}^\nu]_{\dE} &= -\frac{4}{45}\Rf^{\mu\nu} E_{;\mu\nu} - \frac{1}{10} \Rf_{;\mu}E^{;\mu} - \frac{4}{5} E_{;\mu}{}^\mu{}_\nu{}^\nu
\end{align}
\begin{align}
[b_4] &= \frac{1}{72}\Rf^2 - \frac{1}{180}\Rf_{\mu\nu}\Rf^{\mu\nu} + \frac{1}{180} \Rf_{\mu\nu\rho\sigma}\Rf^{\mu\nu\rho\sigma} + \frac{1}{30}\nabla^2 \Rf + \frac{1}{6}E_{;\mu}{}^\mu \nn \\
[b_{4;\mu}]_{\dE} &= -\frac{1}{12} \Rf E_{;\mu} + \frac{1}{12}E^{;\nu}{}_{\nu\mu} \nn \\
[b_{4;\mu}{}^\mu]_{\dE} &= \frac{1}{30}\Rf^{\mu\nu}E_{;\mu\nu} - \frac{1}{9}\Rf E_{;\mu}{}^\mu - \frac{1}{15} \Rf^{;\mu} E_{;\mu} + \frac{1}{20}E_{;\mu}{}^\mu{}_\nu{}^\nu + \frac{1}{4} E_{;\mu}E^{;\mu} \nn \\
[b_{4;\mu\nu}]_{\dEtw} &= \frac{1}{4}E_{;\mu} E_{;\nu} \nn \\
[b_{4;\mu\nu\rho}]_{\dEtw} &= \frac{1}{3}(E_{;\mu} E_{;\nu\rho} + E_{;\nu} E_{;\rho\mu} + E_{;\rho} E_{;\mu\nu}) \nn \\
[b_{4;\mu}{}^\mu{}_\nu{}^\nu]_{\dEtw} &= \frac{5}{6}E^{;\mu}E_{;\mu\nu}{}^\nu + \frac{2}{3}E^{;\mu}E^{;\nu}{}_{\nu\mu} + \frac{4}{9} (E_{;\mu}{}^\mu)^2 + \frac{8}{9}E_{;\mu\nu}E^{;\mu\nu}
\end{align}
\begin{align}
[b_6]_{\dE} &=  \frac{1}{90}\Rf^{\mu\nu}E_{;\mu\nu} + \frac{1}{36}\Rf E_{;\mu}{}^\mu + \frac{1}{30} \Rf^{;\mu} E_{;\mu} + \frac{1}{60}E_{;\mu}{}^\mu{}_\nu{}^\nu + \frac{1}{12} E_{;\mu}E^{;\mu} \nn\\
[b_{6;\mu}]_{\dEtw} &= \frac{1}{12}(E_{;\nu} E^{;\nu}{}_{;\mu} - E_{;\mu} E^{;\nu}{}_{;\nu}) \nn \\
[b_{6;\mu}{}^\mu]_{\dEtw} &= \frac{1}{24}\Rf E_{;\mu}E^{;\mu} - \frac{1}{9}(E_{;\mu}{}^\mu)^2 + \frac{2}{45}E_{;\mu\nu}E^{;\mu\nu} + \frac{1}{15}E^{;\mu}E_{;\mu\nu}{}^\nu - \frac{1}{10}E^{;\mu}E^{;\nu}{}_{\nu\mu}     \nn\\
[b_{6;\mu\nu\rho}]_{\dEth} &= -\frac{1}{8} E_{;\mu} E_{;\nu} E_{;\rho}\nn \\
[b_{6;\mu}{}^{\mu}{}_\nu{}^\nu]_{\dEth} &= - \frac{2}{3}E^{;\mu} E^{;\nu} E_{;\mu\nu} - \frac{1}{3} E_{;\mu}E^{;\mu}E_{;\nu}{}^{;\nu}
\end{align}
\begin{align}
[b_8]_{\dEtw} &=\frac{1}{72}\Rf E_{;\mu}E^{;\mu} + \frac{1}{72}(E_{;\mu}{}^\mu)^2 + \frac{1}{90}E_{;\mu\nu}E^{;\mu\nu} + \frac{1}{60}E^{;\mu}E_{;\mu\nu}{}^\nu + \frac{1}{60}E_{;\mu}E_{;\nu}{}^{\nu\mu}  \nn \\
[b_{8;\mu}]_{\dEth} &= -\frac{1}{24} E_{;\nu} E^{;\nu} E_{;\mu}  \nn \\
[b_{8;\mu}{}^{;\mu}]_{\dEth} &= -\frac{1}{12} E_{;\mu} E_{;\nu} E^{;\mu\nu} - \frac{1}{72} E_{;\mu}E^{;\mu}E_{;\nu}{}^{;\nu} \nn\\
[b_{8;\mu\nu\rho\sigma}]_{\dEfo} &= \frac{1}{16}E_{;\mu} E_{;\nu} E_{;\rho} E_{;\sigma}
\end{align}
\begin{align}
[b_{10}]_{\dEth} &= \frac{1}{60}E_{;\mu\nu}E^{;\mu}E^{;\nu} + \frac{1}{72}E_{;\mu}{}^{\mu}E^{;\nu}E_{;\nu} \nn\\
[b_{10;\mu\nu}]_{\dEfo} &= \frac{1}{48} E_{;\rho} E^{;\rho} E_{;\mu} E_{;\nu}
\end{align}
\begin{align}
[b_{12}]_{\dEfo} &= \frac{1}{12\cdot 4!}(E^{;\mu}E_{;\mu})^2
\end{align}

Using these relations we can then write the final result for the heat kernel at coincident points as
\begin{equation}
K(x,x;\tk) = (4\pi\tk)^{-(\df+1)/2}\sum_{k=0} b_{2k}\tk^k e^{\tk E},
\end{equation}
where the relevant heat kernel coefficients are given by
\begin{align}\label{eq:HKcoeffsapp}
[b_0] &= 1, \nn \\
[b_2] &= \frac{1}{6}\Rf,\nn\\
\lbrack b_4\rbrack &= \frac{1}{72}\Rf^2 - \frac{1}{180}\Rf_{\mu\nu}\Rf^{\mu\nu} + \frac{1}{180} \Rf_{\mu\nu\rho\sigma}\Rf^{\mu\nu\rho\sigma} + \frac{1}{30}\nabla^2 \Rf + \frac{1}{6}E_{;\mu}{}^\mu,  \nn\\
\lbrack b_6 \rbrack_{\dE} &=  \frac{1}{90}\Rf^{\mu\nu}E_{;\mu\nu} + \frac{1}{36}\Rf E_{;\mu}{}^\mu + \frac{1}{30} \Rf^{;\mu} E_{;\mu} + \frac{1}{60}E_{;\mu}{}^\mu{}_\nu{}^\nu + \frac{1}{12} E_{;\mu}E^{;\mu},\nn \\
[b_8]_{\dEtw} &=\frac{1}{72}\Rf E_{;\mu}E^{;\mu} + \frac{1}{72}(E_{;\mu}{}^\mu)^2 + \frac{1}{90}E_{;\mu\nu}E^{;\mu\nu} + \frac{1}{60}E^{;\mu}E_{;\mu\nu}{}^\nu + \frac{1}{60}E_{;\mu}E_{;\nu}{}^{\nu\mu},\nn \\
[b_{10}]_{\dEth} &= \frac{1}{60}E_{;\mu\nu}E^{;\mu}E^{;\nu} + \frac{1}{72}E_{;\mu}{}^{\mu}E^{;\nu}E_{;\nu}, \nn\\
[b_{12}]_{\dEfo} &= \frac{1}{12\cdot 4!}(E^{;\mu}E_{;\mu})^2.
\end{align}
As a cross-check, we can compare our results for the coefficients above with the results of \cite{Vassilevich:2003xt,vandeVen:1997pf}. To see this, one must first expand the exponential $e^{\tk E}$ as a power series in $\tk$.  Doing so, and carefully accounting for cross-terms between the expansion of $e^{\tk E}$ and the heat kernel expansion itself, one recovers the results of \cite{Vassilevich:2003xt} for the $[b_0],$ $[b_2],$ $[b_4]$ and $[b_6]_{\partial E}$ terms. The terms $[b_8]_{\partial E^2},$ $[b_{10}]_{\partial E^3}$ and $[b_{12}]_{\partial E^4}$, on the other hand, can be matched with the flat space results of \cite{vandeVen:1997pf}. The only term that this procedure does not account for is the term in $[b_8]_{\partial E^2}$ which includes the Ricci scalar $\Rf$ and which has not (to our knowledge) previously been computed. Interestingly, this term (along with several others) can be seen to appear if one assumes the lower-order heat kernel coefficients each individually exponentiate. It would be interesting to understand this apparent exponentiation property further.

% ----------------------------------------------------------------------------

\subsection{Reproducing the standard heat kernel expansion}\label{subsec:HKRecov}

We now turn to verifying that the fixed-$k$ heat kernel expression derived above integrates to the usual form of the heat kernel expansion. In particular, we perform the integral over $k$ for asymptotically AdS black branes, with metrics of the form (\ref{eq:AdSSMetric}) with $f(r)=\frac{r^2}{L^2}(1-\frac{r_h^d}{r^d})$ and $d\Omega_{d-1}^2$ replaced with $L^{-2} d\vec{x}^2.$ This means that in this subsection we specialize to the geometry in (\ref{eq:2Dslice}) with $\df{+}1 =2,$ which is relevant for $(d{+}1)$-dimensional AdS Schwarzschild black branes.  

% .............................................................................

\subsubsection{Leading heat kernel coefficients for black branes}

We will first verify the leading order heat kernel coefficients for the generic $(d{+}1)$-dimensional AdS-Schwarzschild black brane. This will include all of the divergences for scalar fields in the BTZ black hole. The heat kernel (including the appropriate measure factors) expressed as an integral of the fixed-$k$ heat kernel is given by
\begin{equation}\label{eq:HKb2genD}
r^{d-1} \, K(x,x;\tk) = \frac{1}{4\pi\tk}\int \frac{d^{d-1}k}{(2\pi)^{d-1}}\, e^{-\tk(k^2 X +m^2)}\left(1 + \frac{\tk}{6}\left(\Rtw + 6\tilde{E} - \tk k^2 X_{;\mu}{}^\mu + \frac{1}{2}\tk^2 k^4 X_{;\mu}X^{;\mu}\right)\right),
\end{equation}
where we have defined $E = \tilde{E} - k^2 X$ with
\begin{eqnarray}\label{eq:EXapp}
\tilde{E} &=&  -\frac{d^2-1}{4L^2} - \frac{(d-1)^2}{4L^2}\frac{r_h^d}{r^d} \,,\nn\\
%- \frac{(d-3)(d-1)f(r)}{4r^2} - \frac{2(d-1)f'(r)}{4r}, \nn \\
X&=&\frac{1}{r^2} \,,
\end{eqnarray}
and we have expanded the $k$-independent term $\tilde E$ of the exponential as a power series in $\tk$.  The factor of $r^{d-1}$ on the left-hand side of
(\ref{eq:HKb2genD}) arises from%
\footnote{
  Dimensional analysis is
  the simplest way to get straight whether there should have
  also been any overall factors of $L$ in (\ref{eq:HKb2genD}).
  Note that we are keeping here the convention of the main text that
  ${\vec k}$ is dimensionless.  That is, ${\vec k} = {\vec p}L$ here,
  where ${\vec p}$ is the momentum conjugate
  to the transverse position ${\vec x}$.
}
the different volume factors
$\sqrt{g}$ and $\sqrt{g_{(2)}}$ in
\begin {equation}
   \ln Z
   = \frac12 \int d^{d+1}x \sqrt{g} \int \frac{d\tk}{\tk} \, K(x,x;\tk)
   = \frac12 \int \frac{d^{d-1}k}{(2\pi)^{d-1}} \int d^2x \sqrt{g_{(2)}} \,
        \int \frac{d\tk}{\tk} \, K_k(x,x;\tk) .
\end {equation}
In what follows, recall that the curvature invariants and covariant derivatives in (\ref{eq:HKb2genD}) are defined with respect to the two-dimensional geometry (\ref{eq:2Dslice}).

The relevant momentum integrals are given by
\begin{eqnarray}
\int \frac{d^{d-1}k}{(2\pi)^{d-1}} \,  k^{2n}\, e^{-\frac{\tk k^2}{r^2}} &=& \frac{\text{Vol}(\mathbb{S}^{d-2})}{(2\pi )^{d-1}}\int_0^\infty dk \, k^{d-2+2n} \,e^{-\frac{\tk k^2}{r^2}} \nn \\
&=& (4\pi\tk)^{-(d-1)/2} \tk^{-n} r^{d-1+2n}\,\frac{\Gamma(n + \frac{d-1}{2})}{\Gamma(\frac{d-1}{2})} \,.
\end{eqnarray}
Evaluating (\ref{eq:HKb2genD}) we find
\begin{equation}
K(x,x;\tk) = \frac{e^{-\tk m^2}}{(4\pi\tk)^{(d+1)/2}}\left(1 + \frac{\tk}{6}\left(-\frac{d(d+1)}{L^2}\right)\right),
\end{equation}
which is the correct expression for the leading terms of the heat kernel expansion in the AdS-Schwarzschild black brane in $d+1$ dimensions, given that the $(d{+}1)$-dimensional Ricci curvature is $R=-d(d+1)/L^2.$

% .............................................................................

\subsubsection{Including the $a_4$ coefficient}

We now move on to the $O(\tk^2)$ coefficient in the heat kernel expansion, called $a_4$ in the notation of equation (\ref{eq:HKnormal}). This is given by a sum of terms from the coefficients $b_{4}$ to $b_{12}$. Including this term in the heat kernel gives
\begin{align}\label{eq:HKb4genD}
r^{d-1} \, K(x,x;\tk) =  \frac{1}{4\pi\tk} &
\int \frac{d^{d-1}k}{(2\pi)^{d-1}}\,e^{-\tk(k^2 X +m^2)} \nn \\ \times
\Bigg(&1 + \frac{\tk}{6}\left(\Rtw + 6\tilde{E} - \tk k^2 X_{;\mu}{}^\mu + \frac{1}{2}\tk^2 k^4 X_{;\mu}X^{;\mu}\right) + \frac{\tk^2}{6}\left(\Rtw\tilde E + 3\tilde E^2\right)\nn \\
&+\frac{\tk^2}{360}\left(5 \Rtw^2 -2\Rtw_{\mu\nu}\Rtw^{\mu\nu} + 2 \Rtw_{\mu\nu\rho\sigma}\Rtw^{\mu\nu\rho\sigma} + 12 \Rtw_{;\mu}{}^\mu + 60 \tilde{E}_{;\mu}{}^\mu\right) \nn \\
& -\frac{k^2\tk^3}{180}\Big(\Rtw^{\mu\nu}X_{;\mu\nu}+5\Rtw X_{;\mu}{}^\mu + 6 \Rtw^{;\mu}X_{;\mu} \nn \\
& \qquad \qquad \qquad + 6 X_{;\mu}{}^\mu{}_\nu{}^\nu + 30 \tilde{E}^{;\mu} X_{;\mu} + 30 \tilde{E} X_{;\mu}{}^\mu\Big) \nn \\
& + \frac{k^4\tk^4}{360}\Big(5 \Rtw X_{;\mu}X^{;\mu} + 5\left(X_{;\mu}{}^\mu\right)^2 + 4X_{;\mu\nu}X^{;\mu\nu} \nn \\
&\qquad\qquad\qquad + 6 X^{;\mu}X_{\mu\nu}{}^\nu + 6X^{;\mu}X_{\nu}{}^{\nu\mu} + 30 \tilde{E} X_{;\mu}X^{;\mu}\Big) \nn \\
& - \frac{k^6\tk^5}{360}\left(6X_{;\mu\nu}X^{;\mu}X^{;\nu} + 5 X_{;\mu}{}^\mu X_{;\nu}X^{;\nu}\right) + \frac{k^8t^6}{12\cdot4!}\left(X^{;\mu}X_{;\mu}\right)^2\Bigg),
\end{align}
where $\tilde{E}$ and $X$ are given in (\ref{eq:EXapp}) and again all curvature invariants and covariant derivatives are defined with respect to the two-dimensional geometry (\ref{eq:2Dslice}).

After evaluating the various terms, performing the momentum integrals
we find the final expression
\begin{align}\label{eq:HKgenD}
K(x,x;\tk) = \,& (4\pi\tk)^{-(d+1)/2}e^{-\tk m^2} \nn \\
&\times \Bigg(1 + \frac{\tk}{6}\left(-\frac{d(d+1)}{L^2}\right) + \frac{\tk^2}{360 L^4}\bigg(d(d+1)(5d^2+3d+4) + 2d(d-1)^2(d-2)\frac{r_h^{2d}}{r^{2d}}\bigg)\Bigg).
\end{align}
One can verify that this is the appropriate expression for a minimally coupled scalar in the AdS$_{d+1}$ black brane by evaluating the curvature invariants for the metric (\ref{eq:AdSSMetric}) with flat horizon.  These are
\begin{align}
R_{;\mu}{}^\mu &=0, \nn \\
R^2 &= \frac{d^2(d+1)^2}{L^4}, \nn\\
R_{\mu\nu}R^{\mu\nu} &= \frac{d^2(d+1)}{L^4}, \nn \\
R_{\mu\nu\rho\sigma}R^{\mu\nu\rho\sigma} &= 2\frac{d(d+1)}{L^4} + d(d-1)^2(d-2)\frac{r_h^{2d}}{r^{2d}L^4}.
\end{align}
Evaluating the known heat kernel coefficient
\begin{equation}
a_4 = \frac{1}{360}\left(5 R^2 - 2R_{\mu\nu}R^{\mu\nu} + 2 R_{\mu\nu\rho\sigma}R^{\mu\nu\rho\sigma} + 12 R_{;\mu}{}^\mu\right)
\end{equation}
[which is (\ref{eq:HKacoeff4})
with $E$ in that context set to zero to obtain the
$(d{+}1)$-dimensional calculation of $\ln\det(-\nabla^2+m^2)$],
we find precise agreement with (\ref{eq:HKgenD}).

% ===========================================================================

\section{Large-\boldmath$k$ expansion vs.\ large \boldmath$r$}
\label{app:largishr}

\subsection{Potential breakdown of large-$k$ expansion (\ref{eq:questionable})}

In the main text, we reported that one of the terms in our
large-$k$ heat kernel expansion $F(k)$ comes from large enough $r$
to cast into doubt the usefulness of the expansion.  Here we provide
a little more detail about the power counting.  As in
(\ref{eq:HKBTZintgrtdLogZk}), start from
\begin {equation}
F(k) \equiv \ln Z_k^{\rm trunc} = \frac{1}{2}\int_{r_h}^{r_b} dr\int_0^{1/T} d\tau \int_{\Lambda^{-2}}^\infty \frac{d\tk}{\tk}\, K^{\rm trunc}_k(x;\tk)
\equiv
\frac{1}{2T}\int_{r_h}^{r_b} dr \> {\cal K}(r) ,
\end {equation}
and use the formula (\ref{eq:HKBTZtint}) for
what we define here as
\begin {equation}
  {\cal K}(r) \equiv \int \frac{d\tk}{\tk} \, K^{\rm trunc}_k(x;\tk) .
\end {equation}
The IR divergence (\ref{eq:IRdiv}) comes from the fact that
(\ref{eq:HKBTZtint}) approaches a constant, ${\cal K}(\infty)$,
as $r \to \infty$.
The IR divergence is uninteresting; so let's isolate it from our discussion
by subtracting it away, focusing on the IR-regulated contribution
\begin {equation}
  F_{\rm reg}(k) \equiv
  \frac{1}{2T}\int_{r_h}^{\infty} dr
  \bigl[ {\cal K}(r) - {\cal K}(\infty) \bigr] .
\end {equation}

Let's now focus on a particular term in (\ref{eq:HKBTZtint}):
\begin {equation}
\int_{\Lambda^{-2}}^\infty \frac{d\tk}{\tk} K_k^{\rm trunc}(x;\tk) ~=~
\cdots +
  \frac{k^2 + \hat m^2 r^2}{4\pi r^2}
    \ln\left(\frac{k^2+\hat m^2r^2}{\Lambda^2r^2}\right)
  + \cdots ,
\label {eq:calKterm}
\end{equation}
which is one of the terms generated from integrating the leading
``1'' term in
the expansion (\ref{eq:BTZfixedkHK}) of $K_k$.  The corresponding
contribution to $F_{\rm reg}(k)$ above is
\begin {equation}
  \frac{1}{2T}\int_{r_h}^{\infty} dr
  \left[
    \frac{k^2 + \hat m^2 r^2}{4\pi r^2}
      \ln\left(\frac{k^2+\hat m^2r^2}{\Lambda^2r^2}\right)
    - \frac{\hat m^2}{4\pi} \ln\left( \frac{\hat m^2}{\Lambda^2}\right)
  \right] . 
\label {eq:FregTerm}
\end{equation}
There are two important scales in this integral for large $k$:
the scale $r \sim r_h$
set by the integration limit, and the scale $r \sim k/\hat m$
characteristic of the integrand.  The contribution to (\ref{eq:FregTerm})
from $r \sim r_h$ will be of order $r_h/T$ times the integrand evaluated
at $r \sim r_h$, and so of order $k^2/r_h T \sim \hat k^2$ (times
a logarithm) for large $k$.  In contrast, the contribution
to (\ref{eq:FregTerm})
from $r \sim k/\hat m$ will be of order $r/T$
times the integrand evaluated at that $r$,
and so of order $k \hat m/T \sim \hat k \hat m$.
This is the origin of the contribution (\ref{eq:questionable}) discussed
in the main text.  In fact, the entire
$\hat k \hat m L \operatorname{arctan}(\hat k/\hat m L)$ term in
(\ref{eq:HKBTZintgrtdLogZk}) comes from the integral
(\ref{eq:FregTerm}), which gives
\begin {equation}
\frac{1}{4}\bigg[- \hat k^2 \ln(\Lambda^2L^2)  - 2\hat k^2 + \left(\hat k^2 - \hat m^2L^2 \right) \ln(\hat k^2 + \hat m^2L^2) + 4 \hat k \hat m L \arctan\bigg(\frac{\hat k}{\hat m L}\bigg)
+2 \hat m^2 L^2 \ln(\hat m^2 L^2)
\bigg] .
\end {equation}
If one looks at the {\it other} terms in (\ref{eq:calKterm}), there are
also individual contributions of order $k\hat m$ from $r \sim k/\hat m$,
but these all cancel among those other terms, leaving only the
$k\hat m$ contribution that comes from (\ref{eq:FregTerm}).

The moral of the story is that there are large-$r$ contributions
from $r \sim k/m$ that generate the contribution (\ref{eq:questionable})
to $F(k)$.  The large-$k$ expansion that we truncated to determine
$F(k)$ assumed that the expansion parameter $\tk_{\rm eff}$ described in
section \ref{sec:kHK} was small.  Recall for the discussion in that
section that the exponential
$\exp\bigl(-\tk(m^2-E)\bigr) \simeq \exp\bigl(-\tk(m^2+k^2/r^2)\bigr)$ in
(\ref{eq:fixedkHKexp}), or equivalently
$\exp\bigl(-\tk(k^2X+m^2)\bigr)$ in (\ref{eq:BTZfixedkHK}), forces
$t \lesssim r^2/k^2$.  For the largest $\tk$, which is $\tk \sim r^2/k^2$,
the expansion in $t_{\rm eff}$, which is given by (\ref{eq:BTZfixedkHK})
in the BTZ case, is an expansion in $\tk/L^2 \sim r^2/k^2 L^2$.
For $r$ as large as $r \sim k/\hat m$, this is then an expansion in
$\tk/L^2 \sim (\hat m L)^{-2}$.  Unless $\hat m L$ is large (which we
do not want to generally assume in our problem), this expansion parameter
is not small at those large values of $r$.

% ---------------------------------------------------------------------------

\subsection {Using AdS for large $r$}

At large $r$, the spacetime is approximately AdS.  AdS is simple enough
that we do not have to resort to the fixed-$k$ heat kernel
{\it expansion} in powers of $\tk$; we may instead
directly compute the exact result for fixed $k$ (equivalent to summing up the
expansion to all orders).  In this section, we will see that the
exact result reproduces (\ref{eq:questionable}), and so there was no
problem after all as long as we indeed used $\hat m$ for our calculations
in the main text.  [If we had instead done our fixed-$k$ heat kernel
expansion in the main text in terms of the original $m$ rather than
$\hat m$, we would have found something different than (\ref{eq:questionable})
at the order of our expansion, which would have been (a) wrong, and
(b) not a polynomial in $\Delta$ and so the mistake would not be
absorbable into $\Poltk$.]

% ............................................................................

\subsubsection {Three dimensions}

So let's turn to the analysis in AdS.  We will focus first on the case
of Euclidean AdS$_3$ (also known as $H^3$) relevant to the large $r$ behavior of
BTZ.  In this appendix, we will work in units where $L=1$, and
we work in the parametrization
\be\label{eq:EAdS3}
ds^2=\frac{dz^2+d\tau^2+dx_1^2}{z^2} \,,
\ee
where $z$ corresponds to $L^2/r$ in our earlier notation, and $x_1$
is the transverse spatial coordinate. The full partition function is given in terms of the heat kernel
$K(x,x';\tk)$ as in (\ref{eq:detD}) as
\begin {equation}
  \ln Z = \frac12 \int d^3x \sqrt{g}\int_0^\infty \frac{d\tk}{\tk}\> K(x,x;\tk) .
\label {eq:AdSHK}
\end{equation}
The heat kernel $K(x,x';\tk)$
depends on $(x,x')$ only through the chordal distance
\be\label{eq:chordalEAdS3}
  u(x,x')\equiv\frac{(z-z')^2+(\tau-\tau')^2+(x_1-x_1')^2}{2 z z'}
  \equiv \cosh \xi(x,x')-1
\ee 
and is given by \cite{camporesi,Giombi:2008vd}
\be\label{eq:H3HK}
K(\xi;\tk)=\frac{1}{(4\pi\tk)^{3/2}} \frac{\xi}{\sinh\xi}\exp\bigl(-(m^2+1)\tk-\tfrac{\xi^2}{4\tk}\bigr).
\ee
[For AdS, $\xi^2/2$ is the $\sigma$ of (\ref{eq:sigma}).]

We are interested in a fixed-$k$ heat kernel, where $k$ is the conjugate variable to $x_1$.  We can rewrite (\ref{eq:AdSHK}) as
\begin {equation}
  \ln Z = \sum_k \ln Z_k
\end{equation}
with
\begin {equation}\label{eq:ZkAdS}
  \ln Z_k = \frac12 \int dz \> d\tau \> \sqrt{g}
            \int_0^\infty \frac{d\tk}{\tk}\> 
            \int_{-\infty}^\infty d(\Delta x_1) \> e^{-ik \,\Delta x_1}
            K(z,\tau,x_1;z,\tau,x_1+\Delta x_1;\tk) .
\end {equation}
So we are interested in (i) the coincident case of $z=z'$ and $\tau=\tau'$
but (ii)
the Fourier transform with respect to $\Delta x_1 \equiv x_1'-x_1$.  In this case,
\be
u=\frac{(\Delta x_1)^2}{2z^2}=\cosh(\xi)-1=2\sinh^2(\tfrac{\xi}{2}).
\ee
We trade the integral over $\Delta x_1$ for an integral over
$\xi$ while holding $z$ fixed.
Substituting $\Delta x_1 = 2z\sinh(\xi/2)$ gives
\begin {align}\label{eq:fixedkHKAdS3}
  K_k(z;\tk) &\equiv
  \int_{-\infty}^\infty d(\Delta x_1) \> e^{-ik\,\Delta x_1}
     K(z,\tau,x_1;z,\tau,x_1+\Delta x_1;\tk)
\nn\\
  &=
  \frac{z}{(4\pi\tk)^{3/2}}e^{-(m^2+1)\tk}\int_{-\infty}^{\infty} d\xi \>
     e^{-2 i k z \sinh(\tfrac\xi2)}e^{-\tfrac{\xi^2}{4\tk}}
  \frac{\xi}{2\sinh(\tfrac\xi2)} \,.
\end {align}
Note that the $\xi$ integration limits are $-\infty$ and $\infty$ because
the $x_1$ integration limits are.

Using $\sqrt{g} = z^{-3}$, the $z$ and $\tk$ integrals in (\ref{eq:ZkAdS})
give
\begin {multline}\label{eq:intlimitfirst}
\int_{0}^{\infty} dz \sqrt{g} \int_0^{\infty} \frac{d\tk}{\tk} \, K_k(z;\tk)
\\
=\int_{0}^{\infty} dz\int_0^{\infty} d\tk \int_{-\infty}^{\infty} d\xi
\frac{1}{z^2 \tk (4\pi\tk)^{3/2}}e^{-(m^2+1)\tk}
e^{-2 i k z \sinh(\tfrac\xi2)}e^{-\tfrac{\xi^2}{4\tk}}\frac{\xi}{2\sinh(\tfrac\xi2)} \,.
\end {multline}
Given the behavior of the integrand under $\xi \to -\xi$ and under
$z \to -z$,
we can trade the range $(-\infty,+\infty)$ on the $\xi$ integral for
a range $(-\infty,+\infty)$ on the $z$ integral to rewrite
\begin {multline}\label{eq:intlimitswap}
\int_{0}^{\infty} dz \sqrt{g} \int_0^{\infty} \frac{d\tk}{\tk} \, K_k(z;\tk)
\\
=\int_{-\infty}^{\infty} dz\int_0^{\infty} d\tk \int_0^{\infty} d\xi
\frac{1}{z^2 \tk (4\pi\tk)^{3/2}}e^{-(m^2+1)\tk}
e^{-2 i k z \sinh(\tfrac\xi2)}e^{-\tfrac{\xi^2}{4\tk}}\frac{\xi}{2\sinh(\tfrac\xi2)} \,.
\end {multline}
The $z$ integral is
\be
\int_{-\infty}^{\infty} \frac{dz}{z^2} \, e^{-2 i k z \sinh(\tfrac\xi2)}
= -2 \pi |k| \sinh(\tfrac\xi2) .
\ee
The $\xi$ integral can then be done with
\be
\int_0^\infty d\xi \> e^{-\tfrac{\xi^2}{4\tk}} \xi = 2\tk ,
\ee
and then the remaining $\tk$ integral with
\be
\int_0^{\infty} \frac{d\tk}{(4\pi\tk)^{3/2}} \, e^{-(m^2+1)\tk}
= - \frac{\hat m}{4\pi} ,
\ee
where $\hat m^2 \equiv m^2 + 1$ (for the case $d{=}2$ here)
is the same shifted mass (\ref{eq:mhat})
introduced in the main text.
The final result for the integrals is
\be
\int_{0}^{\infty} dz \sqrt{g} \int_0^{\infty} \frac{d\tk}{\tk} \, K_k(z;\tk)
= \frac12 |k| \hat m ,
\ee
giving
\begin {equation}
  \ln Z_k = \int d\tau \, \frac14 |k| \hat m .
\end {equation}

In the context of thermal AdS or the asymptotic AdS region of BTZ, the integral
over $\tau$ just gives a factor of $1/T$, in which case the above
result becomes
\begin {equation}\label{eq:ZkAdS3}
  \ln Z_k = \frac{|k| \hat m}{4 T} = \frac{\pi}{2} |\hat k| \hat m ,
\end {equation}
using the definition (\ref{eq:khat}) of $\hat k$.
Restoring factors of $L$, this is exactly the same as the term
(\ref{eq:questionable}) identified in the main text, and so
(\ref{eq:questionable}) is correct in spite of the worries one
might have had about the fixed-$k$ heat kernel expansion for this
term.

% ............................................................................

\subsubsection {Generalizing to higher dimensions}

The linear-in-$\hat m$ term above generalizes to higher odd-dimensional cases $H^{d+1}$ (or Euclidean AdS$_{d+1}$) as well. We will see that it has precisely the same form as (\ref{eq:ZkAdS3}), but with $\hat m$ replaced by the
general-$d$ formula (\ref{eq:mhat}) for the shifted mass.

To start, we note a few properties of the geometry, Laplacian and heat kernel in $H^{d+1}.$ The geometry of $H^{d+1}$ is given by replacing the coordinate $x_1$ with a $(d{-}1)$-vector $\vec{x}$ in the metric (\ref{eq:EAdS3}) and similarly in the chordal distance (\ref{eq:chordalEAdS3}). When acting on a function of the geodesic distance $\xi,$ the scalar Laplacian $\Delta_{d+1}$ on $H^{d+1}$ can be written in the simple form
\begin{equation}
\Delta_{d+1} = \partial_\xi^2 + d\,\coth\xi\,\partial_\xi.
\end{equation}
As observed by Camporesi \cite{camporesi}, this implies the following recursion relation between Laplacians in different dimensions:
\begin{equation}\label{eq:LapRR}
\Delta_{d+1} \mathcal D = \mathcal D (\Delta_{d-1} - d + 1),
\end{equation}
where we have defined the operator
\begin{equation}
\mathcal D \equiv \frac{1}{\sinh \xi} \frac{\partial}{\partial\xi}.
\end{equation}
Since the heat kernel in $H^{d+1},$ which we denote here by $K_{d+1}(\xi,\tk),$ satisfies
\begin{equation}
\left(\partial_\tk -  \Delta_{d+1} + m^2\right)K_{d+1}(\xi;\tk) = 0,
\end{equation}
we can use (\ref{eq:LapRR}) to derive the recursion relation 
\begin{equation}\label{eq:HypHKrecurs}
K_{d+1}(\xi;\tk) = -\frac{1}{2\pi}\frac{e^{-(d-1)\tk}}{\sinh\xi}\frac{\partial }{\partial\xi}K_{d-1}(\xi;\tk),
\end{equation}
where the normalization is fixed by demanding the appropriate behavior as $t\rightarrow0.$ This recursion relation will prove useful in evaluating $\ln Z_k$ for $H^{d+1}.$ We also note in passing that the exponential in (\ref{eq:HypHKrecurs}) is precisely that required to shift the $\hat m^2$ in $(d{-}1)$-dimensions to the $\hat m^2$ in $(d{+}1)$-dimensions, as appropriate for the higher dimensional heat kernel such that
\begin{equation}
K_{d+1}(\xi;\tk) \propto e^{-\tk \hat m_{(d+1)}^2},
\end{equation}
with $\hat m_{(d+1)}^2 = m^2 + \frac{d^2}{4} = \left(\Delta - \frac{d}{2}\right)^2.$

To derive $\ln Z_k$ we follow precisely the same steps as in the previous section. In what follows we will explicitly compute $\ln Z_k$ for the physically relevant cases of $d=4$ and $d=6.$ We again define the fixed-$k$ heat kernel by taking the coincidence limits in the $z$ and $\tau$ coordinates as in (\ref{eq:fixedkHKAdS3}) and perform the higher dimensional version of the Fourier transform 
\begin{align}
K^{(d+1)}_{k}(z;\tk) &\equiv \int_{-\infty}^\infty d^{d-1}(\Delta x) \, e^{-i\vec{k}\cdot\Delta\vec{x}} K_{d+1}(\xi;\tk) \nn\\
&= \text{Vol}(S^{d-3}) \int_0^\infty \rho^{d-2} d\rho \int_{0}^{\pi} d\theta (\sin \theta)^{d-3} e^{-i k \rho\cos\theta} K_{d+1}(\rho,z;\tk),
\end{align}
where $\rho = |\Delta\vec{x}|,$ $k=|\vec k|,$ and $\theta$ is the polar angle which we define as the angle between $\vec{k}$ and $\Delta\vec{x}.$ Evaluating the polar integral for $d=4$ and $d=6$ we find  
\begin{eqnarray}
K_k^{(5)}(z;\tk)  
&=& \frac{2\pi i}{k} \int_{-\infty}^{\infty} d\rho\, \rho \, e^{-i k \rho} K_5(\rho,z;\tk),\nn \\
K_k^{(7)}(z;\tk) 
&=& \frac{4\pi^2 i}{k^3} \int_{-\infty}^{\infty}d\rho\, \rho \, (1+i k \rho)e^{-i k \rho} K_7(\rho,z;\tk).
\end{eqnarray}

To compute $\ln Z_k^{(d+1)}$ we should take these expressions and use them to evaluate the integrals
\begin{eqnarray}
\ln Z_k^{(d+1)} &=& \frac{1}{2}\int_0^{1/T} d\tau\int_0^\infty dz \sqrt{g} \int_0^\infty \frac{d\tk}{\tk}K^{(d+1)}_k(z;\tk) \nn \\ 
&=& \frac{1}{2T}\int_0^\infty \frac{dz}{z^{d+1}}  \int_0^\infty \frac{d\tk}{\tk}K^{(d+1)}_k(z;\tk).
\end{eqnarray}
Making the change of variables from $\rho$ to $\xi$ by using $\rho = 2 z\sinh\ft{\xi}{2}$ and focusing on the $z$ and $\xi $ integrals we have 
\begin{eqnarray}\label{eq:HK57}
\int_0^\infty \frac{dz}{z^5} K_k^{(5)}(z;\tk) &=& \frac{2\pi i}{k}\int_0^\infty \frac{dz}{z^3}\int_{-\infty}^{\infty}d\xi \sinh\xi\,e^{-2i k z\sinh(\ft{\xi}{2})} K_5(\xi;\tk), \nn \\
\int_0^\infty \frac{dz}{z^7} K_k^{(7)}(z;\tk) &=& \frac{4\pi^2 i}{k^3}\int_0^\infty \frac{dz}{z^5}\int_{-\infty}^{\infty}d\xi \sinh\xi\,\left(1+2i k z \sinh\ft{\xi}{2}\right)e^{-2ik z\sinh(\ft{\xi}{2})} K_7(\xi;\tk).
\end{eqnarray}
We can now use the recursion relation to replace $K_5(\xi;\tk)$ with $K_3(\xi,\tk)$ and $K_7(\xi;\tk)$ with $K_5(\xi,\tk)$. In particular, 
\begin{eqnarray}
K_5(\xi;\tk) &=& - \frac{1}{2\pi} \frac{e^{-3\tk}}{\sinh\xi}\frac{\partial}{\partial\xi} K_3(\xi;\tk), \nn\\
K_7(\xi;\tk) &=& - \frac{1}{2\pi} \frac{e^{-5\tk}}{\sinh\xi}\frac{\partial}{\partial\xi} K_5(\xi;\tk).
\end{eqnarray}
Inserting these relations in (\ref{eq:HK57}) and integrating by parts we find
\begin{eqnarray}\label{eq:intbypartHK57}
\int_{0}^\infty \frac{dz}{z^5} K_k^{(5)}(z;\tk) &=& e^{-3\tk}\int_0^\infty \frac{dz}{z^3}\int_{-\infty}^{\infty}d\xi z \cosh(\ft{\xi}{2})\,e^{-2i k z\sinh(\ft{\xi}{2})} K_3(\xi;\tk), \nn \\
\int_{0}^\infty \frac{dz}{z^7} K_k^{(7)}(z;\tk) &=& e^{-5\tk}\frac{2\pi i}{k}\int_0^\infty \frac{dz}{z^3}\int_{-\infty}^{\infty}d\xi \sinh\xi\,e^{-2i k z\sinh(\ft{\xi}{2})} K_5(\xi;\tk).
\end{eqnarray}

There are two things to notice here. First, if we substitute (\ref{eq:H3HK}) into the expression on the first line we reproduce precisely the same integral in (\ref{eq:intlimitfirst}) that computed $\ln Z_k^{(3)}$ (the fixed-$k$ partition function in three dimensions), up to a factor of $e^{-3\tk}.$ As mentioned previously, this additional exponential factor is precisely that required to shift the $\hat m_{(3)}^2 = m^2 + 1$ in (\ref{eq:H3HK}) to $\hat m_{(5)}^2 = m^2 + 4$ such that $-\hat m_{(5)}^2 \tk$ appears in the exponential. This all means that $\ln Z_k^{(5)}$ will have the same expression as $\ln Z_k^{(3)}$ and will be given by (\ref{eq:ZkAdS3}), except that the $\hat m$ in (\ref{eq:ZkAdS3}) will be the appropriate expression for five dimensions. The second thing to notice in (\ref{eq:intbypartHK57}) is that a similar relation exists between the $K_k^{(7)}(z;\tk)$ integral on the second line of (\ref{eq:intbypartHK57}) and the $K_k^{(5)}(z;\tk)$ integral on the first line of (\ref{eq:HK57}). Similar reasoning, and applying one additional step of recursion, then implies that $\ln Z_k^{(7)}$ also has the same form as $\ln Z_k^{(3)}.$

In the end, we see that the recursion relation between heat kernels in odd-dimensional $H^{d+1}$ implies a simple relation between the fixed-$k$ partition functions $\ln Z_k^{(d+1)}.$ In particular, we have 
\begin{equation}
\ln Z_k^{(d+1)} = \frac{\pi}{2}|\hat k| \hat m_{(d+1)},
\end{equation}
when $d+1$ is odd. 

The case when $d+1$ is even can be worked out similarly, and one can verify that $\ln Z_k^{(d+1)}$ is linear in $k.$ However, evaluating the explicit dependence on $\hat m$ is more complicated because in this case the recursion relation should reduce the final result to an expression as an integral of the heat kernel on $H^2,$ which is not known to be expressible in terms of elementary functions \cite{camporesi}.

% ===========================================================================

\section{Doing the momentum sums for BTZ}\label{app:PoissonResum}

The sum over momentum modes in equation (\ref{eq:BTZfixedkHK}) can be done analytically using Poisson resummation (see \cite{Dunne} for a related discussion). Consider first the sum on $k$ of the leading term in (\ref{eq:BTZfixedkHK}), which can be rewritten as
\begin{equation}\label{eq:Poisson}
\frac{1}{4\pi\tk}\sum\limits_{k=-\infty}^{\infty} e^{-\tk(k^2 X+\hat m^2)} = \frac{1}{4\pi\tk}\sqrt{\frac{\pi}{X \tk}}\sum_{\ell=-\infty}^{\infty} e^{-\frac{\pi^2\ell^2}{X \tk} - \tk \hat m^2}.
\end{equation}
To find the partition function we need to integrate this over $\tk$ as in (\ref{eq:IntHK}). For $\ell=0$ the integral is divergent, and so we evaluate that term independently. Using a strict UV cutoff $t \geq 1/\Lambda^2$, one finds
\begin{eqnarray}\label{eq:BTZt0l0}
\int_{1/\Lambda^2}^\infty\frac{d\tk}{\tk}\frac{1}{4\pi\tk}\sqrt{\frac{\pi}{X \tk}} e^{-\tk \hat m^2} &=&\frac{1}{4\sqrt{\pi X}}\left(\frac{2}{3}\Lambda^3 - 2\hat m^2 \Lambda + \frac{4}{3}\hat m^3 \sqrt{\pi}\right) + O(\Lambda^{-1}) .
\end{eqnarray}
This is precisely the behavior expected for the UV-divergent contribution arising in standard heat kernel regularization. In particular, substituting $X=1/r^2$, we see that the prefactor is proportional to the BTZ volume element factor $1/\sqrt{X} = \sqrt{g} = r.$

For $\ell\neq0$, the integral on $\tk$ is finite and yields a Bessel function. Using the integral representation
\begin{equation}
K_\alpha(z) = \frac{1}{2}\left(\frac{z}{2}\right)^\alpha \int_0^\infty \frac{d\tk}{\tk}\,\tk^{-\alpha}e^{-\tk-\frac{z^2}{4\tk}},
\end{equation}
and performing the sum on $\ell\neq0,$ we find
\begin{eqnarray}\label{eq:BTZt0sum}
\sum_{\ell\neq0}\int_0^\infty\frac{d\tk}{\tk}\frac{1}{4\pi\tk}\sqrt{\frac{\pi}{X \tk}} e^{-\frac{\pi^2\ell^2}{X \tk} - \tk \hat m^2} &=& \frac{\hat{m}^{3/2} X^{1/4}}{\pi^2}\sum_{\ell=1}^\infty \frac{1}{\ell^{3/2}}\, K_{3/2}\!\left(\frac{2\pi\hat{m}\ell}{\sqrt{X}}\right) \nn \\
&=&  \frac{X}{4\pi^3}\sum_{\ell=1}^\infty \frac{1}{\ell^{3}}\left(1+ y \ell\right)e^{-y\ell}\nn\\
&=& \frac{X}{4\pi^3}\left(\Li_3\left(e^{-y}\right) +y\,\Li_2\left(e^{-y}\right) \right),
\end{eqnarray}
where we have defined $y=\frac{2\pi\hat m}{\sqrt{X}}$, $\Li_n(x)$ are poly-logarithms, and in the second line we have used $K_{3/2}(x) = \sqrt{\frac{\pi}{2}}x^{-3/2}e^{-x}(1+x)$. Finite contributions of this sort will be crucial in comparing with the standard results for the partition function.

A similar analysis for the second term in (\ref{eq:BTZfixedkHK}) can be done. The only difference is that the integrand contains one additional factor of $\tk$. The $\ell=0$ mode is again divergent.  Performing that integral with the same regulator we find
\begin{eqnarray}
\int_{1/\Lambda^2}^\infty\frac{d\tk}{\tk}\frac{1}{4\pi}\sqrt{\frac{\pi}{X \tk}} e^{- \tk \hat m^2} &=& \frac{1}{4\sqrt{\pi X}}\left(2\Lambda - 2\hat m \sqrt{\pi}\right) + O(\Lambda^{-1}),
\end{eqnarray}
which is again the expected UV divergence and local contributions in a heat kernel regularization. The sum over non-zero $\ell$ in this case gives a representation of $K_{1/2}(x) = \sqrt{\frac{\pi}{2}}x^{-1/2}e^{-x}.$ Performing similar manipulations to the previous case, one finds
\begin{eqnarray}
\sum_{\ell\neq0}\int_0^\infty\frac{d\tk}{\tk}\frac{1}{4\pi} \sqrt{\frac{\pi}{X \tk}}\, e^{-\frac{\pi^2\ell^2}{X \tk} - \tk \hat m^2} &=&  \frac{1}{2\pi}\,\Li_1(e^{-y}).
\end{eqnarray}
Putting it together, this yields
\begin{eqnarray}\label{eq:BTZt1}
\int \frac{d\tk}{\tk}\frac{1}{4\pi}\sum\limits_{k=-\infty}^{\infty} e^{-\tk(k^2 X+\hat m^2)} &=& \frac{1}{4\sqrt{\pi X}}\left(2\Lambda - 2\hat m \sqrt{\pi}\right) + \frac{1}{2\pi}\,\Li_1(e^{-y}) + O(\Lambda^{-1}).
\end{eqnarray}
Finally, we need expressions for the two sums with factors of $\tk k^2$ in (\ref{eq:BTZfixedkHK}). These can be determined from the previous result in a simple manner. We can generate the $tk^2$ terms by simply differentiating the previous result with respect to $X,$ namely
\begin{eqnarray}
\int \frac{d\tk}{\tk}\frac{1}{4\pi}\sum\limits_{k=-\infty}^{\infty} \tk k^2 e^{-\tk(k^2 X+\hat m^2)} &=& -\frac{d}{dX}\left( \int \frac{d\tk}{\tk}\frac{1}{4\pi}\sum\limits_{k=-\infty}^{\infty} e^{-\tk(k^2 X+\hat m^2)}\right),\\
\int \frac{d\tk}{\tk}\frac{1}{4\pi}\sum\limits_{k=-\infty}^{\infty} \tk^2 k^4 e^{-\tk (k^2 X+\hat m^2)} &=& \frac{d^2}{dX^2}\left( \int \frac{d\tk}{\tk}\frac{1}{4\pi}\sum\limits_{k=-\infty}^{\infty} e^{-\tk(k^2 X+\hat m^2)}\right).
\end{eqnarray}
Applying these relations to  (\ref{eq:BTZt1}) we find
\begin{align}\label{eq:BTZt2sum}
\int \frac{d\tk}{\tk}\frac{1}{4\pi}\sum\limits_{k=-\infty}^{\infty} \tk k^2 e^{-\tk(k^2 X+\hat m^2)} &= \frac{\Lambda}{4\sqrt{\pi X^3}} - \frac{\hat m}{4\sqrt{X^3}}\left(1 + 2\,\Li_0(e^{-y})\right) + O(\Lambda^{-1}), \\
\label{eq:BTZt3sum}
\int \frac{d\tk}{\tk}\frac{1}{4\pi}\sum\limits_{k=-\infty}^{\infty} \tk^2 k^4 e^{-\tk(k^2 X+\hat m^2)} &= \frac{3\Lambda}{8\sqrt{\pi X^5}} - \frac{3\hat m}{8\sqrt{X^5}}\left(1 + 2\,\Li_0(e^{-y})\right) + \frac{\hat m^2\pi}{2X^3}\,\Li_{-1}(e^{-y}) + O(\Lambda^{-1}). 
\end{align}
Note that many of the poly-logarithms above can be simply expressed in terms of elementary functions.  In particular,
\begin{eqnarray}
\Li_1(e^{-y}) &=& - \ln(1-e^{-y}), \nn \\
\Li_0(e^{-y}) &=& \frac{e^{-y}}{1-e^{-y}}, \nn\\
\Li_{-1}(e^{-y}) &=&  \frac{e^{-y}}{\left(1-e^{-y}\right)^2}.
\end{eqnarray}
Poly-logarithms are convenient, as they satisfy the simple relation
\begin{equation}\label{eq:plog}
\frac{d}{dy} \Li_n(e^{-y}) = - \Li_{n-1}(e^{-y}).
\end{equation}

Taking the expressions from this section and evaluating $I$ by performing the integral over $d^2x = dr\,d\tau$ in (\ref{eq:IntHK}) with $X = 1/r^2$ as in (\ref{eq:X}), one can derive the result for $I$ quoted in (\ref{eq:HKsummed}). Interestingly, as we will see below, the $r$ dependence of the integral ends up as a total derivative and the final result is easily expressed in terms of temperature dependent poly-logarithms plus a local integral containing the UV divergent terms.

Putting together (\ref{eq:BTZt0l0} and (\ref{eq:BTZt0sum}) and integrating we find
\begin{eqnarray}\label{eq:BTZt0fin}
\frac{1}{2}\int_{r_h}^{\infty} dr \int_0^{1/T} d\tau \int_{1/\Lambda^2}^\infty\frac{d\tk}{\tk}\frac{1}{4\pi\tk}\sum\limits_{k=-\infty}^{\infty} e^{-\tk(k^2/r^2+\hat m^2)} &=& \int d^3x \sqrt{g}\left(\frac{\Lambda^3}{24\pi^{3/2}} - \frac{\Lambda \hat m^2}{8\pi^{3/2}} + \frac{\hat m^3}{12\pi}\right) \nn \\
&&  + \frac{1}{(2\pi)^2}\frac{1}{(2\pi TL)^2}\, \Li_3 (e^{-4\pi^2 \hat m T L^2}),
\end{eqnarray}
where $r_h = 2\pi L^2 T$ and we have replaced $2\pi = \int d\phi$ in order to write the measure in the first line. In the second line we have used that, for $X = 1/r^2$, the final line of (\ref{eq:BTZt0sum}) is a total derivative
\begin{equation}
\frac{1}{r^2}\left[\Li_3\left(e^{-2\pi \hat m r}\right) +2\pi \hat m r\,\Li_2\left(e^{-2\pi \hat m r}\right) \right] = -\left(\frac{1}{r}\,\Li_3\left(e^{-2\pi \hat m r}\right)\right)'
\end{equation}
and $\lim\limits_{r\rightarrow\infty}\frac{1}{r} \Li_3\left(e^{-2\pi \hat m r}\right) = 0.$

Next, using (\ref{eq:BTZt1}), (\ref{eq:BTZt2sum}), and (\ref{eq:BTZt3sum}) we can evaluate the remaining terms in (\ref{eq:BTZfixedkHK}),
\begin{align}\label{eq:BTZt1fin}
\frac{1}{2}\int_{r_h}^{\infty} & dr \int_0^{1/T} d\tau \int_{1/\Lambda^2}^\infty\frac{d\tk}{\tk}\frac{1}{4\pi\tk}\sum\limits_{k=-\infty}^{\infty} e^{-\tk(k^2 X + \hat m^2)}\frac{\tk}{6}\left(\Rtw + 6\tilde{E} + \frac{3d^2}{2L^2} - \tk k^2 X_{;\mu}{}^\mu + \frac{1}{2}\tk^2 k^4 X_{;\mu}X^{;\mu}\right) \nn\\
&= -\frac{1}{2T}\int_{r_h}^\infty dr \bigg[\frac{r}{24\pi L^2}\Big( \Li_1(e^{-y}) +y\,\Li_0(e^{-y})\Big) - \frac{r_h^2\hat m}{4\pi L^2}\left(\frac{1}{y}\,\Li_1(e^{-y}) + \frac{1}{3}\,\Li_0(e^{-y})\right)\bigg]' \nn \\
&= -\frac{r_h}{24\pi TL^2}\,\Li_1\left(e^{-2\pi \hat m r_h}\right) \nn \\
&= -\frac{1}{12}\,\Li_1\left(e^{-4\pi^2 \hat mTL^2}\right),
\end{align}
where $y=2\pi \hat m r$  and where a prime denotes a derivative with respect to $r.$ In deriving the second line of (\ref{eq:BTZt1fin}) we have used the relation (\ref{eq:plog}). Finally, it is worth noting that with $\hat m$ as the mass in the exponential all of the $\Lambda$ dependence (as well as all non-vanishing terms in the large mass limit) sits in the first term in the heat kernel expansion, which was evaluated in (\ref{eq:BTZt0fin}).
 
%%%%%%%%%%%%%%%%%%%%%%%%%%%%%%%%%%%%%%%%%%%%%%%%%%%%%%%%%%%%%%%%%%%%%%%%%%%%%%

%%%%%%%%%%%%%%%%%%%%%%%%%%%%%%%%%%%%%%%%%%%%%%%%%%%%%%%%%%%%%%%%%%%%%%%%%%%%%%

\begin{thebibliography}{}

\bibitem{Denef:2009kn}
F.~Denef, S.~A.~Hartnoll and S.~Sachdev,
``Black hole determinants and quasinormal modes,''
Class.\ Quant.\ Grav.\  {\bf 27}, 125001 (2010)
[arXiv:0908.2657 [hep-th]].

\bibitem{Gibbons:1994cg}
G.~W.~Gibbons and S.~W.~Hawking,
``Euclidean quantum gravity,''
Singapore, Singapore: World Scientific (1993) 586 p.

\bibitem{GKT}
  S.~S.~Gubser, I.~R.~Klebanov and A.~A.~Tseytlin,
  ``Coupling constant dependence in the thermodynamics
    of N=4 supersymmetric Yang-Mills theory,''
  Nucl.\ Phys.\ B {\bf 534}, 202 (1998)
  [hep-th/9805156].

\bibitem{MPS}
  R.~C.~Myers, M.~F.~Paulos and A.~Sinha,
  ``Quantum corrections to $\eta/s$,''
  Phys.\ Rev.\ D {\bf 79}, 041901 (2009)
  [arXiv:0806.2156 [hep-th]].
 

\bibitem{Kovtun:2003vj}
P.~Kovtun and L.~G.~Yaffe,
``Hydrodynamic fluctuations, long time tails, and supersymmetry,''
Phys.\ Rev.\ D {\bf 68}, 025007 (2003)
[hep-th/0303010].

\bibitem{CaronHuot:2009iq}
S.~Caron-Huot and O.~Saremi,
``Hydrodynamic Long-Time tails From Anti de Sitter Space,''
JHEP {\bf 1011}, 013 (2010)
[arXiv:0909.4525 [hep-th]].

\bibitem{Denef:2009yy}
F.~Denef, S.~A.~Hartnoll and S.~Sachdev,
``Quantum oscillations and black hole ringing,''
Phys.\ Rev.\ D {\bf 80}, 126016 (2009)
[arXiv:0908.1788 [hep-th]].

\bibitem{Hartnoll:2009kk}
S.~A.~Hartnoll and D.~M.~Hofman,
``Generalized Lifshitz-Kosevich scaling at quantum criticality from the holographic correspondence,''
Phys.\ Rev.\ B {\bf 81}, 155125 (2010)
[arXiv:0912.0008 [cond-mat.str-el]].

\bibitem{Anninos:2010sq}
D.~Anninos, S.~A.~Hartnoll and N.~Iqbal,
``Holography and the Coleman-Mermin-Wagner theorem,''
Phys.\ Rev.\ D {\bf 82}, 066008 (2010)
[arXiv:1005.1973 [hep-th]].

\bibitem{Faulkner:2010da}
T.~Faulkner, N.~Iqbal, H.~Liu, J.~McGreevy and D.~Vegh,
``From Black Holes to Strange Metals,''
arXiv:1003.1728 [hep-th].

\bibitem{Faulkner:2013bna}
T.~Faulkner, N.~Iqbal, H.~Liu, J.~McGreevy and D.~Vegh,
``Charge transport by holographic Fermi surfaces,''
Phys.\ Rev.\ D {\bf 88}, 045016 (2013)
[arXiv:1306.6396 [hep-th]].

\bibitem{Hartnoll:2010gu}
S.~A.~Hartnoll and A.~Tavanfar,
``Electron stars for holographic metallic criticality,''
Phys.\ Rev.\ D {\bf 83}, 046003 (2011)
[arXiv:1008.2828 [hep-th]].

\bibitem{Allais:2012ye}
A.~Allais, J.~McGreevy and S.~J.~Suh,
``A quantum electron star,''
Phys.\ Rev.\ Lett.\  {\bf 108}, 231602 (2012)
[arXiv:1202.5308 [hep-th]].

\bibitem{Allais:2013lha}
A.~Allais and J.~McGreevy,
``How to construct a gravitating quantum electron star,''
Phys.\ Rev.\ D {\bf 88}, no. 6, 066006 (2013)
[arXiv:1306.6075 [hep-th]].

\bibitem{Datta:2011za} 
S.~Datta and J.~R.~David,
``Higher Spin Quasinormal Modes and One-Loop Determinants in the BTZ black Hole,''
JHEP {\bf 1203}, 079 (2012)
%doi:10.1007/JHEP03(2012)079
[arXiv:1112.4619 [hep-th]].

\bibitem{Datta:2012gc} 
S.~Datta and J.~R.~David,
``Higher spin fermions in the BTZ black hole,''
JHEP {\bf 1207}, 079 (2012)
%doi:10.1007/JHEP07(2012)079
[arXiv:1202.5831 [hep-th]].

\bibitem{Zhang:2012kya} 
H.~b.~Zhang and X.~Zhang,
``One loop partition function from normal modes for $ \mathcal{N}=1$ supergravity in AdS$_3$,''
Class.\ Quant.\ Grav.\  {\bf 29}, 145013 (2012)
%doi:10.1088/0264-9381/29/14/145013
[arXiv:1205.3681 [hep-th]].

\bibitem{Zojer:2012rj} 
T.~Zojer,
``On gravity one-loop partition functions of three-dimensional critical gravities,''
Class.\ Quant.\ Grav.\  {\bf 30}, 075005 (2013)
%doi:10.1088/0264-9381/30/7/075005
[arXiv:1210.6887 [hep-th]].

\bibitem{Keeler:2014hba}
C.~Keeler and G.~S.~Ng,
``Partition Functions in Even Dimensional AdS via Quasinormal Mode Methods,''
JHEP {\bf 1406}, 099 (2014)
[arXiv:1401.7016 [hep-th]].

\bibitem{Keeler:2016wko} 
C.~Keeler, P.~Lisbao and G.~S.~Ng,
``Partition Functions with spin in $AdS_2$ via Quasinormal Mode Methods,''
arXiv:1601.04720 [hep-th].

\bibitem{Maloney:2016gsg} 
A.~Maloney and S.~F.~Ross,
``Holography on Non-Orientable Surfaces,''
arXiv:1603.04426 [hep-th].

\bibitem{Warnick:2013hba}
  C.~M.~Warnick,
  ``On quasinormal modes of asymptotically anti-de Sitter black holes,''
  arXiv:1306.5760 [gr-qc].

\bibitem{Coleman}
 S.~R.~Coleman,
 ``The uses of instantons,''
  in Aspects of symmetry, Cambridge University Press, UK (1988).

\bibitem{Natario:2004jd}
J.~Natario and R.~Schiappa,
``On the classification of asymptotic quasinormal frequencies for d-dimensional black holes and quantum gravity,''
Adv.\ Theor.\ Math.\ Phys.\  {\bf 8}, 1001 (2004)
[hep-th/0411267].

\bibitem{Musiri:2005ev}
S.~Musiri, S.~Ness and G.~Siopsis,
``Perturbative calculation of quasi-normal modes of AdS Schwarzschild black holes,''
Phys.\ Rev.\ D {\bf 73}, 064001 (2006)
[hep-th/0511113].

\bibitem{Arnold:2013gka}
P.~Arnold and P.~Szepietowski,
``Spin 1/2 quasinormal mode frequencies in Schwarzschild-AdS spacetime,''
Phys.\ Rev.\ D {\bf 88}, 086002 (2013)
[arXiv:1308.0341 [hep-th]].

\bibitem{Arnold:2013zva}
P.~Arnold, P.~Szepietowski and D.~Vaman,
``Gravitino and other spin-3/2 quasinormal modes in Schwarzschild-AdS spacetime,''
Phys.\ Rev.\ D {\bf 89}, no. 4, 046001 (2014)
[arXiv:1311.6409 [hep-th]].

\bibitem{Siopsis:2008xz}
G.~Siopsis,
``Analytic calculation of quasi-normal modes,''
Lect.\ Notes Phys.\  {\bf 769}, 471 (2009)
[arXiv:0804.2713 [hep-th]].

\bibitem{Vassilevich:2003xt}
D.~V.~Vassilevich,
 ``Heat kernel expansion: User's manual,''
 Phys.\ Rept.\  {\bf 388}, 279 (2003)
 [hep-th/0306138].

\bibitem{Mann:1996ze}
R.~B.~Mann and S.~N.~Solodukhin,
``Quantum scalar field on three-dimensional (BTZ) black hole instanton: Heat kernel, effective action and thermodynamics,''
Phys.\ Rev.\ D {\bf 55}, 3622 (1997)
[hep-th/9609085].
 
\bibitem{Festuccia:2008zx} 
 G.~Festuccia and H.~Liu,
 ``A Bohr-Sommerfeld quantization formula for quasinormal frequencies of AdS black holes,''
 Adv.\ Sci.\ Lett.\  {\bf 2}, 221 (2009)
 [arXiv:0811.1033 [gr-qc]].

\bibitem{Breitenlohner:1982jf}
P.~Breitenlohner and D.~Z.~Freedman,
``Stability in Gauged Extended Supergravity,''
Annals Phys.\  {\bf 144}, 249 (1982).

\bibitem{DeWitt:1965jb}
B.~S.~DeWitt,
``Dynamical theory of groups and fields,''
Conf.\ Proc.\ C {\bf 630701}, 585 (1964)
[Les Houches Lect.\ Notes {\bf 13}, 585 (1964)].

\bibitem{Poisson:2011nh}
E.~Poisson, A.~Pound and I.~Vega,
``The Motion of point particles in curved spacetime,''
Living Rev.\ Rel.\  {\bf 14}, 7 (2011)
[arXiv:1102.0529 [gr-qc]].

\bibitem{Barvinsky:1985an} 
A.~O.~Barvinsky and G.~A.~Vilkovisky,
``The Generalized Schwinger-Dewitt Technique in Gauge Theories and Quantum Gravity,''
Phys.\ Rept.\  {\bf 119}, 1 (1985).

\bibitem{vandeVen:1997pf}
A.~E.~M.~van de Ven,
``Index free heat kernel coefficients,''
Class.\ Quant.\ Grav.\  {\bf 15}, 2311 (1998)
[hep-th/9708152].

\bibitem{camporesi} 
R.~Camporesi,
``Harmonic analysis and propagators on homogeneous spaces,''
Phys.\ Rept.\  {\bf 196}, 1 (1990).
%doi:10.1016/0370-1573(90)90120-Q

\bibitem{Giombi:2008vd} 
S.~Giombi, A.~Maloney and X.~Yin,
``One-loop Partition Functions of 3D Gravity,''
JHEP {\bf 0808}, 007 (2008)
%doi:10.1088/1126-6708/2008/08/007
[arXiv:0804.1773 [hep-th]].

\bibitem{Dunne}
G.~V.~Dunne,
``Functional determinants in quantum field theory,''
Lecture notes given at ``the 14th WE Heraeus Saalburg summer school,''
Wolfersdorf, Thuringia, September 2008. http://www.itp.uni-hannover.de/saalburg/Lectures/dunne.pdf


\end{thebibliography}
\end{document}